\documentclass[12pt]{article}
\usepackage{amssymb, amsmath, bm, float, epsfig, color, slashed, dsfont, subcaption, multirow, arydshln}
\usepackage{graphicx}
\usepackage[centering]{geometry}
\usepackage{cite}
\usepackage{color}
\allowdisplaybreaks[4]

\makeatletter
\@addtoreset{equation}{section}
\makeatother

\begin{document}

\thispagestyle{empty}

\begin{flushright}
EPHOU-22-019\\
KYUSHU-HET-249\\
KOBE-TH-22-05 \\
\end{flushright}
\vspace{40pt}
\begin{center}
{\Large\bf {Index theorem on magnetized blow-up manifold of $T^2/\mathbb{Z}_N$}
} \\

\vspace{40pt}
{\bf{Tatsuo Kobayashi$^{\ast}\hspace{.5pt}$}\footnote{E-mail: kobayashi@particle.sci.hokudai.ac.jp}}, \,
{\bf{Hajime Otsuka$^{\ast\ast}\hspace{.5pt}$}\footnote{E-mail: otsuka.hajime@phys.kyushu-u.ac.jp}}, \,
{\bf{Makoto Sakamoto$^\dagger\hspace{.5pt}$}\footnote{E-mail: dragon@kobe-u.ac.jp}}, \, 
{\bf{Maki Takeuchi$^\dagger\hspace{.5pt}$}\footnote{E-mail: 191s107s@stu.kobe-u.ac.jp}}, \, 
{\bf{Yoshiyuki Tatsuta$^{\dagger\dagger}\hspace{.5pt}$}\footnote{E-mail: yoshiyuki.tatsuta@sns.it}}, \,
{\bf{Hikaru Uchida$^{\ast}\hspace{.5pt}$}\footnote{E-mail: h-uchida@particle.sci.hokudai.ac.jp}} \\

\vspace{40pt}
\it $^{\ast}$ Department of Physics, Hokkaido University, Sapporo 060-0810, Japan \\[5pt]
{\it $^{\ast\ast}$ Department of Physics, Kyusyu University, 744 Motooka, Nishi-ku, Fukuoka, 819-0395, Japan \\[5pt]
\it $^\dagger$ Department of Physics, Kobe University, Kobe 657-8501, Japan \\[5pt]
$^{\dagger\dagger}$ Scuola Normale Superiore and INFN, Piazza dei Cavalieri 7, 56126 Pisa, Italy} \\
\end{center}
\vspace{30pt}
\begin{abstract}
	\noindent 
We investigate blow-up manifolds of $T^2/{\mathbb{Z}}_N\,(N=2,3,4,6)$ orbifolds with magnetic flux $M$. Since the blow-up manifolds have no singularities, we can apply the Atiyah-Singer index theorem to them. Then, we establish the zero-mode counting formula $n_{+}-n_{-}=(M-V_{+})/N+1$, where $V_{+}$ denotes the sum of  winding numbers at fixed points on the $T^2/{\mathbb{Z}}_N$ orbifolds, as the Atiyah-Singer index theorem on the orbifolds, and clarify physical and geometrical meanings of the formula.
\end{abstract}

\newpage
\setcounter{page}{2}
\setcounter{footnote}{0}

\section{Introduction}
The Atiyah-Singer index theorem~\cite{Atiyah:1963zz} states that the index of a Dirac operator $\slashed{D}$ 
\begin{align}
{\rm{Ind}}(i \slashed {D}) \equiv n_{+} -n_{-}
\label{Indexop}
\end{align}
is a topological invariant. Here, $n_{\pm}$ are the numbers of $\pm$ chiral zero modes for the Dirac operator. The index theorem has been applied to many areas in physics, such as the chiral anomaly in gauge theory~\cite{PhysRevLett.42.1195,PhysRevD.22.1499}, the Witten index~\cite{Witten:1982df}, and anomaly inflow~\cite{Witten:2019bou,Ivanov:2022jor}.

\color{black}
In particular, we are interested in counting the number of chiral zero modes appearing in the four-dimensional (4d) effective field theories. The Standard Model has a lot of mysteries unanswered, including the generation problem of the quarks and leptons and the fermion mass hierarchy, and also to naturally explain their flavor structure. String theory and higher dimensional theory are strong candidates beyond the Standard Model. Many proposals have been made to solve the generation problem, but known mechanisms to produce degenerate chiral zero modes are very limited. It is known to obtain the chiral spectra as magnetic flux compactifications in type-I and II string theory~\cite{Abouelsaood:1986gd,Blumenhagen:2000wh,Angelantonj:2000hi,Angelantonj:2002ct,Blumenhagen:2005mu,Blumenhagen:2006ci,Ibanez:2012zz} and heterotic string theory \cite{Anderson:2011ns,Anderson:2012yf,Abe:2015mua,Otsuka:2018oyf}. These models have provided semi-realistic models of string phenomenology, e.g. three generation models~\cite{Abe:2008sx,Abe:2015yva}, fermion mass hierarchy~\cite{Cremades:2004wa}, and flavor structure~\cite{Abe:2014vza,PhysRevD.94.035031,Kobayashi:2016qag,Buchmuller:2017vho,PhysRevD.97.075019,Kikuchi:2021yog,Hoshiya:2022qvr}. 

The Atiyah-Singer (AS) index theorem for a two-dimensional (2d) compact manifold ${\mathcal{M}}^2$ with magnetic flux is known as~\cite{Witten:1984dg,Green:1987mn}
\begin{align}
n_{+}-n_{-}=\frac{1}{2\pi}\int_{{\mathcal{M}}^2}F,
\label{Indintro}
\end{align}
where $F$ is a 2-form field strength of the flux. We should here stress that the index $n_{+}-n_{-}$ is determined only by the flux but not the curvature of the 2d manifold ${\mathcal{M}}^2$. A simple application of the index theorem \eqref{Indintro} is to take ${\mathcal{M}}^2$ to be a 2d torus $T^2$:
\begin{align}
n_{+}-n_{-}=\frac{1}{2\pi}\int_{T^2}F=M,
\label{IndTorusintro}
\end{align}
where $M$ is an integer and corresponds to a magnetic flux quantization number.

The application of the index theorem to $T^2/{\mathbb{Z}}_N\,\,(N=2,3,4,6)$ magnetized orbifolds will be phenomenologically and mathematically interesting because the index $n_{+}-n_{-}$ gives the generation number in 4d effective field theories and it turns out to complicatedly depend on the flux quanta $M$, the ${\mathbb{Z}}_N$ eigenvalues under the ${\mathbb{Z}}_N$ transformation, the Scherk-Schwarz (SS) twist phases $(\alpha_1,\alpha_2)$ and $N$ (see the last columns in Tables $1-5$ of the appendix)~\cite{Abe:2008fi,Abe:2013bca,Abe:2014noa,PhysRevD.96.096011,Sakamoto:2020pev,PhysRevD.103.025009}. In Ref.\cite{Sakamoto:2020pev}, a complete list of the index has been shown to satisfy the zero-mode counting formula\footnote{By use of the trace formula, Eq.\eqref{Zeromodecountingintro} has been derived for $M=0$ in Ref.~\cite{PhysRevD.103.025009} and for arbitrary $M$ with $N=2$ in Ref.\cite{Imai}.}
\begin{align}
n_{+}-n_{-}=\frac{M}{N}-\frac{V_+}{N}+1,
\label{Zeromodecountingintro}
\end{align}
where $V_{+}$ is the sum of the winding numbers at the fixed points of $T^2/{\mathbb{Z}}_N$ orbifolds.

The first term on the right-hand side of Eq.\eqref{Zeromodecountingintro} could be understood from Eq.\eqref{IndTorusintro} because the area of the $T^2/{\mathbb{Z}}_N$ orbifolds reduces to $1/N$ of that of the torus $T^2$. The origin of the second and third terms on the right-hand side of Eq.\eqref{Zeromodecountingintro} is, however, unclear, and those terms seem not to be related to any flux on the $T^2/{\mathbb{Z}}_N$ orbifolds. In fact, Eq.\eqref{Zeromodecountingintro} has not been derived as the AS index theorem in Ref.~\cite{Sakamoto:2020pev} and the relation \eqref{Zeromodecountingintro} has been verified by computing the values of $n_{+}-n_{-}$ and $M/N-V_{+}/N+1$, separately and then by simply comparing them.

Our main purposes of this paper are to understand the formula \eqref{Zeromodecountingintro} as the AS index theorem and clarify physical and geometrical meanings of the formula. There is, however, a problem: The $T^2/{\mathbb{Z}}_N$ orbifolds have singularities, and the AS index theorem cannot directly be applied to singular ``manifolds". Our strategy to overcome the problem is to construct smooth blow-up manifolds of the $T^2/{\mathbb{Z}}_N$ orbifolds by removing cones around the singularities of the $T^2/{\mathbb{Z}}_N$ orbifolds and replacing them with parts of the 2d sphere $S^2$~\cite{Kobayashi:2019fma,Kobayashi:2019gyl}. Then, we can apply the AS index theorem directly to the blow-up manifolds. From the blow-up procedure, we can confirm the formula \eqref{Zeromodecountingintro} as the AS index theorem and clarify the physical and geometrical meanings of the second and the third terms on the right-hand side of Eq.\eqref{Zeromodecountingintro}.

This paper is organized as follows: In Section 2, we briefly review zero modes on the $T^2/{\mathbb{Z}}_N\,\,(N=2,3,4,6)$ orbifolds. In Section 3, we construct the blow-up manifolds and obtain two important conditions.
In Section 4, we derive the AS index theorem on the blow-up manifolds and reinterpret the zero-mode counting formula. Section 5 is devoted to the conclusion. In the appendix, we show the detailed results of Section 5. 

\section{Magnetized $T^2/{\mathbb{Z}}_N$ orbifold}

\subsection{Magnetized $T^2$}
We review the $U(1)$ gauge theory on a 2d torus with homogeneous magnetic flux~\cite{Cremades:2004wa}. 
First of all, let us consider the six-dimensional (6d)  space-time, which contains {4d} Minkowski space-time ${\cal{M}}^4$ and an extra {2d} torus $T^2$ with magnetic flux. 
The Lagrangian of a 6d Weyl fermion in magnetic flux background is given by
\begin{align}
{\mathcal{L}}_{6d}=i\bar{\Psi}\Gamma^{M} D_{M} \Psi, \qquad \Gamma_7 \Psi = \Psi,
 \end{align}
 where $M(=0,1,2,3,5,6)$ is the 6d spacetime index and $D_M=\partial_M-iqA_M$ is the covariant derivative. $\Gamma^M$ is 6d gamma matrix and $\Gamma_7$ denotes the 6d chirality operator.

By the Kaluza-Klein mode expansion, the 6d Weyl fermion $\Psi(x,z)$ can be decomposed into 4d Weyl left/right-handed fermions 
$\psi_{{\rm{L}}/{\rm{R}}}^{(4)}(x)$ as 
\begin{align}
	\Psi(x, z) = \sum_{n, \hspace{.5pt} j} 
	\bigl( \psi^{(4)}_{\textrm{R}, \hspace{.5pt} n, \hspace{.5pt} j}(x) \otimes \psi^{(2)}_{+, \hspace{.5pt} n, \hspace{.5pt} j}(z) + \psi^{(4)}_{\textrm{L}, \hspace{.5pt} n, \hspace{.5pt} j}(x) \otimes \psi^{(2)}_{-, \hspace{.5pt} n, \hspace{.5pt} j}(z) \bigr),
\end{align}
 where $x^{\mu}(\mu=0,1,2,3)$ denotes the 4d Minkowski coordinate and $z$ is the complex coordinate on $T^2$. The 2d Weyl fermions $\psi^{(2)}_{\pm, \hspace{.5pt} n, \hspace{.5pt} j}(z)$ are  expressed as the form
 \begin{gather}
	\psi^{(2)}_{+, \hspace{.5pt} n, \hspace{.5pt} j}(z) = 
	\begin{pmatrix}
		\psi_{+, \hspace{.5pt} n, \hspace{.5pt} j}(z) \\[3pt] 0
	\end{pmatrix}
	, \qquad \psi^{(2)}_{-, \hspace{.5pt} n, \hspace{.5pt} j}(z) = 
	\begin{pmatrix}
		0 \\[3pt] \psi_{-, \hspace{.5pt} n, \hspace{.5pt} j}(z)
	\end{pmatrix},
	\label{psipm}
\end{gather}
where $n$ and $j$ label the Landau level and the degeneracy of mode functions on each level, respectively.
The torus is defined by the identification $z\sim z+1\sim z+\tau \, (\tau \in {\mathbb{C}},\,\, {\rm{Im}}\tau>0\,)$ with the complex coordinate $z\equiv y_1+\tau y_2$, where ${\bm{y}}=(y_1,y_2)\,\,(0\leq y_1,y_2<1)$ is the oblique coordinate.

\color{black}
The non-zero magnetic flux $f$ on the torus can be obtained as $f=\int_{T^2} F$ with the field strength
 \begin{align}
 F(z)=\frac{if}{2{\rm{Im}}\tau}dz \wedge d\bar{z}.
\end{align}
 For $F=dA$, the (1-form) vector potential is
 \begin{align}
 A(z;\zeta)=\frac{f}{2{\rm{Im}}\tau}{\rm{Im}}((\bar{z}+\bar{\zeta}) dz)\equiv A_z(z;\zeta)dz+A_{\bar{z}}(z;\zeta)d{\bar{z}},
 \end{align}
 and $A_z(z;\zeta)$, $A_{\bar{z}}(z;\zeta)$ are explicitly given by
 \begin{align}
 A_z(z;\zeta)=-\frac{i}{2}\frac{\pi M}{{\rm{Im}}\tau} (\bar{z}+\bar{\zeta}), \qquad
 A_{\bar{z}}(z;\zeta)=\frac{i}{2}\frac{\pi M}{{\rm{Im}}\tau}(z+\zeta),
 \end{align}
where $\zeta \equiv \zeta_1+\tau\zeta_2$ ($\zeta_1,\ \zeta_2 \in \mathbb{R}$) denotes the Wilson line.
Then, we obtain
 \begin{align}
& A(z+1;\zeta)=A(z;\zeta)+d\left( \frac{f}{2{\rm{Im}}\tau}{\rm{Im}}(z+\zeta) \right) \equiv A(z;\zeta)+d\Lambda_1(z+\zeta),\label{gaugetr1}\\
& A(z+\tau;\zeta)=A(z;\zeta)+d\left( \frac{f}{2{\rm{Im}}\tau}{\rm{Im}}(\bar{\tau} (z+\zeta)) \right)\equiv A(z;\zeta)+d\Lambda_2(z+\zeta),\label{gaugetr2}
 \end{align}
where $\Lambda_1(z+\zeta)$ and $\Lambda_2(z+\zeta)$ are gauge parameters. 
It follows from Eqs.\eqref{gaugetr1} and \eqref{gaugetr2} that the torus lattice shifts can be reinterpreted as gauge transformations. 

 \color{black}
The 2d Weyl fermions are required to satisfy the pseudo periodic boundary conditions (BCs)
\begin{align}
\psi_{\pm, \hspace{.5pt} n, \hspace{.5pt} j}(z + 1;\zeta) = U_1(z) \psi_{\pm, \hspace{.5pt} n, \hspace{.5pt} j}(z;\zeta), \qquad \psi_{\pm, \hspace{.5pt} n, \hspace{.5pt} j}(z + \tau;\zeta) = U_2(z) \psi_{\pm, \hspace{.5pt} n, \hspace{.5pt} j}(z;\zeta),
\label{BCs}
\end{align}
with
\begin{align}
U_i(z) = e^{i \Lambda_i (z+\zeta)} e^{2\pi i \alpha_i} \quad (i=1,2),
\label{BCss}
\end{align}
where $\alpha_i \,\, (i=1,2)$ are called SS twist phases which are allowed to be any real numbers. The consistency condition with the contractible loop, $z\to z+1 \to z+1+\tau \to z+\tau \to z $, leads to the magnetic flux quantization condition
\begin{align}
\frac{f}{2\pi}\equiv M \in \mathbb{Z}.
\end{align}

The 2d Weyl fermions satisfy the equations
\begin{align}
-2D_z \psi_{-, \hspace{.5pt} n, \hspace{.5pt} j}(z;\zeta)&=-2(\partial_z -iA_z(z;\zeta))\psi_{-, \hspace{.5pt} n, \hspace{.5pt} j}(z;\zeta)=m_n \psi_{+, \hspace{.5pt} n, \hspace{.5pt} j}(z;\zeta), \label{MD1}\\
2D_{\bar{z}} \psi_{+, \hspace{.5pt} n, \hspace{.5pt} j}(z;\zeta)&=2(\partial_{\bar{z}} -iA_{\bar{z}}(z;\zeta))\psi_{+, \hspace{.5pt} n, \hspace{.5pt} j}(z;\zeta)=m_n \psi_{-, \hspace{.5pt} n, \hspace{.5pt} j}(z;\zeta) .
\label{MD2}
\end{align}
We focus on zero modes with $m_n=0$. From Eqs.\eqref{MD1} and \eqref{MD2}, zero modes satisfy 
 \begin{align}
\left(\partial_z -\frac{\pi M}{2{\rm{Im}}\tau}(\bar{z}+\bar{\zeta})\right)\psi_{-, \hspace{.5pt} 0, \hspace{.5pt} j}(z;\zeta)=0,\qquad
\left(\partial_{\bar{z}} +\frac{\pi M}{2{\rm{Im}}\tau}{(z+\zeta)}\right)\psi_{+, \hspace{.5pt} 0, \hspace{.5pt} j}(z;\zeta)=0.
\label{Zero}
\end{align}
In the case of $M>0$, only $\psi_{+, \hspace{.5pt} 0, \hspace{.5pt} j}$ has the normalizable solutions that satisfy the pseudo periodic BCs \eqref{BCs} and  they are given as
 \color{black}
 \begin{align}
 &\psi_{+,0}^{(j+\alpha_1,\alpha_2)}(z;\zeta) 
 =e^{-\frac{\pi M}{2{\rm{Im}}\tau}|z+\zeta|^2}g^{(j+\alpha_1,\alpha_2)}(z;\zeta)\qquad(j=0,1,\cdots,M-1), \\
 &g^{(j+\alpha_1,\alpha_2)}(z;\zeta)= \mathcal{N}_{T^2}\,e^{\frac{\pi M}{2{\rm{Im}}\tau}(z+\zeta)^2} \, e^{2\pi i\frac{j+\alpha_1}{M}(\alpha_2-M\zeta_1)}\,\vartheta
\begin{bmatrix}
\tfrac{j + \alpha_1}M \\[3pt] -\alpha_2
\end{bmatrix}
(M(z+\zeta), M\tau).
\label{toruses}
\end{align}
Here, $j=0,1,\cdots,|M|-1$ stand for the degeneracy of zero mode solutions, and ${\cal{N}}_{T^2}$ is  a normalization constant. The Jacobi $\vartheta$-function is defined by
\begin{align}
\vartheta
\begin{bmatrix}
a \\[3pt]b
\end{bmatrix}
(z, \tau)
=\sum_{l=-\infty}^{\infty}e^{i\pi(a+l)^2 \tau} e^{2\pi i(a+l)(z+b)}.
\end{align}
Note that the Wilson line $\zeta=\zeta_1+\tau\zeta_2$ can be pushed on the SS phases as
\begin{align}
\alpha_1 \rightarrow \alpha'_1=\alpha_1+ M\zeta_2, \quad \alpha_2 \rightarrow \alpha'_2=\alpha_2-M\zeta_1, 
\label{reSS}
\end{align}
by the $U(1)$ local and gauge transformation:
\begin{align}
\psi_{+,0}^{(j+\alpha_1,\alpha_2)}(z;\zeta) \rightarrow V^{-1}_{\zeta}(z) \psi_{+,0}^{(j+\alpha_1,\alpha_2)}(z;\zeta) &= \psi_{+,0}^{(j+\alpha_1^{\prime},\alpha_2^{\prime})}(z;0),
\label{localU(1)} \\
A(z;\zeta) \rightarrow A(z;\zeta) + i V^{-1}_{\zeta}(z) d V_{\zeta}(z) &= A(z;0),
\label{U(1)gauge}
\end{align}
with
\begin{align}
V^{-1}_{\zeta}(z) \equiv e^{-\pi iM \frac{{\rm Im}(\bar{\zeta}z)}{{\rm Im}\tau}-\pi iM\zeta_1\zeta_2},
\label{U(1)gauge}
\end{align}
as shown in Ref.~\cite{Abe:2013bca}. Hence, hereafter, we set $\zeta=0$.
On the other hand, in the case of $M<0$, only $\psi_{-, \hspace{.5pt} 0, \hspace{.5pt} j}$ has the normalizable solutions, and they are given in a similar way.

The above results are consistent with the 
AS index theorem on the torus with magnetic flux,~i.e. 
\begin{align}
n_{+}-n_{-}=\frac{1}{2\pi}\int_{T^2}F=M.
\label{IndTorus}
\end{align}
The index theorem \eqref{IndTorus} shows that the number of the independent chiral zero modes is decided by the magnetic flux quantization number $M$ on the magnetized torus and further that the generation number of this model is given by $M$. We emphasize that the index $n_{+}-n_{-}$ depends only on the flux.

\subsection{Magnetized $T^2/{\mathbb{Z}}_N$}
In this subsection, we review the $U(1)$ gauge theory on twisted orbifolds $T^2/{\mathbb{Z}}_N$ with magnetic flux~\cite{Abe:2013bca,Abe:2014noa}. It has been known that there are only four kinds of the $T^2/{\mathbb{Z}}_N$ orbifolds with $N=2,3,4,6$. The $T^2/{\mathbb{Z}}_N$ orbifolds are defined by the torus identification and the additional ${\mathbb{Z}}_N$ one
\begin{align}
z\sim \rho z \qquad
(\rho=e^{2\pi i/N}\quad \,(N=2,3,4,6)).
\end{align}
 For $N=2$, there is no restriction on $\tau$ except for ${\rm{Im}}\tau>0$.
 On the other hand, for $N=3,4,6$, $\tau$ should be fixed at $\tau = \rho$ due to the analysis of crystallography.
 
An important feature of the $T^2/{\mathbb{Z}}_N$ orbifolds is the existence of the fixed points $z^{\rm{fp}}_I$ defined by
 \begin{align}
 z^{\rm{fp}}_I=\rho z^{\rm{fp}}_I+ u+v\tau \qquad {\rm{for}} \quad ^{\exists} \,u,v \in \,{\mathbb{Z}}.
 \label{FP}
 \end{align}
The ${\mathbb{Z}}_N$ fixed points on the $T^2/{\mathbb{Z}}_N$ orbifolds are given by
\begin{align}
z^{\rm{fp}}_I=
\begin{cases}
0, \frac{1}{2},\frac{\tau}{2},\frac{1+\tau}{2} & \quad  ~{\rm{on}}\,\, T^2/{\mathbb{Z}}_2,\\ 
0, \frac{2+\tau}{3}, \frac{1+2\tau}{3}& \quad  ~{\rm{on}} \,\,T^2/{\mathbb{Z}}_3,\\ 
0, \frac{1+\tau}{2} & \quad  ~{\rm{on}}\,\,T^2/{\mathbb{Z}}_4,\\
0 & \quad  ~{\rm{on}}\,\, T^2/{\mathbb{Z}}_6,
\end{cases}
\end{align}
 and the respective values of $(u,v)$ in Eq.\eqref{FP} are
 \begin{align}
(u,v)=
\begin{cases}
(0,0), (1, 0), (0, 1), (1, 1) & \quad  ~{\rm{on}}\,\, T^2/{\mathbb{Z}}_2,\\ 
(0,0), (1, 0), (1, 1) & \quad  ~{\rm{on}} \,\,T^2/{\mathbb{Z}}_3,\\ 
(0,0), (1, 0) & \quad  ~{\rm{on}}\,\,T^2/{\mathbb{Z}}_4,\\
(0,0) & \quad  ~{\rm{on}}\,\, T^2/{\mathbb{Z}}_6.
\end{cases}
\end{align}
Note that there are additional fixed points for $N=4,6,$ since the ${\mathbb{Z}}_4\,({\mathbb{Z}}_6)$ group includes ${\mathbb{Z}}_2\,\,({\mathbb{Z}}_2$ and ${\mathbb{Z}}_3)$ as its subgroup. They are not invariant under the ${\mathbb{Z}}_4\,({\mathbb{Z}}_6)$ transforamation, but invariant  under the ${\mathbb{Z}}_2\,\,({\mathbb{Z}}_2$ and ${\mathbb{Z}}_3)$ transformation up to the torus shifts. The additional fixed points are found as 
\begin{align}
{\mathbb{Z}}_2 \,\,{\rm{fixed\,\, points}}:\quad & z^{\rm{fp}}_I=\frac{1}{2},\,\frac{\tau}{2} \qquad\qquad{\rm{on}}\,\, T^2/{\mathbb{Z}}_4,\\
{\mathbb{Z}}_3\,\, {\rm{fixed \,\,points}}:\quad & z^{\rm{fp}}_I=\frac{1+\tau}{3},\,\frac{2+2\tau}{3}\qquad{\rm{on}}\,\, T^2/{\mathbb{Z}}_6,\\
{\mathbb{Z}}_2 \,\,{\rm{fixed \,\,points}}:\quad & z^{\rm{fp}}_I=\frac{1}{2},\,\frac{\tau}{2},\,\frac{1+\tau}{2} \qquad {\rm{on}}\,\, T^2/{\mathbb{Z}}_6.
\end{align}
We should emphasize that the fixed points are singular points on the $T^2/{\mathbb{Z}}_N$ orbifolds.

For the orbifold identification, the Scherk-Schwarz phases $(\alpha_1,\alpha_2)$ must be quantized such as
\begin{gather}
(\alpha_{1}, \alpha_{2})
= (0,0), (1/2,0), (0,1/2), (1/2,1/2) \qquad  {\rm{on}}\,\,~ T^{2}/{\mathbb{Z}}_{2},
\label{Z2_SSphase}  \\
\alpha = \alpha_{1} = \alpha_{2} = 
\begin{cases} 
0, 1/3, 2/3 & \quad (M = \textrm{even})\\
1/6, 3/6, 5/6 & \quad (M = \textrm{odd})
\end{cases} \qquad  {\rm{on}}\,\,~ T^2/{\mathbb{Z}}_3,
\label{Z3_SSphase} \\
\alpha = \alpha_{1} = \alpha_{2}
= 0, 1/2 \qquad  {\rm{on}}\,\,~ T^{2}/{\mathbb{Z}}_{4},
\label{Z4_SSphase} \\
\alpha = \alpha_{1} = \alpha_{2} = 
\begin{cases} 
0 & \quad (M = \textrm{even})\\
1/2 & \quad (M = \textrm{odd})
\end{cases} \qquad {\rm{on}}\,\, ~ T^2/{\mathbb{Z}}_6.
\label{Z6_SSphase}
\end{gather}

Let us discuss ${\mathbb{Z}}_N$ eigenfunctions on the $T^2/{\mathbb{Z}}_N$ orbifolds with magnetic flux. They should obey the boundary conditions \eqref{BCs} and the orbifold boundary conditions
\begin{align}
\psi_{T^2/{\mathbb{Z}}_N^m,+}^{(j+\alpha_1,\alpha_2)}(\rho z)&=\rho^m \psi_{T^2/{\mathbb{Z}}_N^m,+}^{(j+\alpha_1,\alpha_2)}(z), \label{OBC1}\\
\psi_{T^2/{\mathbb{Z}}_N^m,-}^{(j+\alpha_1,\alpha_2)}(\rho z)&=\rho^{m+1} \psi_{T^2/{\mathbb{Z}}_N^m,-}^{(j+\alpha_1,\alpha_2)}(z),
\label{OBC}
\end{align}
where $\rho^m\,(m=0,1,\cdots,N-1)$ in Eq.\eqref{OBC1} denotes the ${\mathbb{Z}}_N$ eigenvalue. 
If the ${\mathbb{Z}}_N$ eigenvalue of $\psi_{T^2/{\mathbb{Z}}_N^m,+}^{(j+\alpha_1,\alpha_2)}(z)$ is $\rho^m$, then that of $ \psi_{T^2/{\mathbb{Z}}_N^m,-}^{(j+\alpha_1,\alpha_2)}(z)$ has to be $\rho^{m+1}$. The difference in eigenvalues  comes from a rotation matrix acting on 2d spinors, and it can also be understood from the relations \eqref{MD1} and \eqref{MD2}. 
Then, the ${\mathbb{Z}}_N$ eigenfunctions can be constructed by the following linear combinations of the wave functions on the torus
\begin{align}
\psi_{T^2/{\mathbb{Z}}_N^m,+}^{(j+\alpha_1,\alpha_2)}(z)&={\mathcal{N}}_{T^2/{\mathbb{Z}}_N,+}
\sum_{k=0}^{N-1} \rho^{-km} \psi_{T^2,+}^{(j+\alpha_1,\alpha_2)}(\rho^k z),\label{MFZN+} \\
\psi_{T^2/{\mathbb{Z}}_N^m,-}^{(j+\alpha_1,\alpha_2)}(z)&={\mathcal{N}}_{T^2/{\mathbb{Z}}_N,-}
\sum_{k=0}^{N-1} \rho^{-k(m+1)} \psi_{T^2,-}^{(j+\alpha_1,\alpha_2)}(\rho^k z),
\label{MFZN}
\end{align}
where ${\mathcal{N}}_{T^2/{\mathbb{Z}}_N,\pm}$ are normalization constants. Especially, zero modes with the ${\mathbb{Z}}_N$ eigenvalue $\rho^m$ are given by
\begin{align}
&\psi_{T^2/{\mathbb{Z}}_N^m,+,0}^{(j+\alpha_1,\alpha_2)}(z)=e^{-\frac{\pi M}{2{\rm{Im}}\tau}|z|^2 }h_{+,m}^{(j+\alpha_1,\alpha_2)}(z), \label{Zeromode+1}\\
&h_{+,m}^{(j+\alpha_1,\alpha_2)}(z)={\mathcal{N}}_{T^2/{\mathbb{Z}}_N,+}\sum_{k=0}^{N-1} \rho^{-km} g^{(j+\alpha_1,\alpha_2)}(\rho^k z).
\label{Zeromode+}
\end{align}
Here, $h_{+,m}^{(j+\alpha_1,\alpha_2)}(z)$ denotes the holomorphic function of $z$. 

Let us investigate the ${\mathbb{Z}}_N$ eigenfunctions $\psi_{T^2/{\mathbb{Z}}_N^m,+}^{(j+\alpha_1,\alpha_2)}(z)$ around the fixed points $ z^{\rm{fp}}_I\equiv y^{\rm{fp}}_{1I}+\tau y^{\rm{fp}}_{2I}$ by modifying Eq.\eqref{OBC1}. Their property will become important later.
First, we define the coordinate $Z$ such that $Z=0$ at the fixed point $z^{\rm{fp}}_I$,~i.e. $Z \equiv z - z^{\rm{fp}}_I$.
Next, we rewrite $z$ by $Z$ as $z = (z - z^{\rm{fp}}_I) + z^{\rm{fp}}_I = Z + z^{\rm{fp}}_I$.
This means that the second term, $z^{\rm{fp}}_I$, can be regarded as the Wilson line $\zeta=z^{\rm{fp}}_I$ ($\zeta_1=y^{\rm{fp}}_{1I}$, $\zeta_2=y^{\rm{fp}}_{2I}$) from the {viewpoint of} the coordinate $Z$. (See the previous subsection.)
Then, the Wilson line can be pushed on SS phases by the $U(1)$ local and gauge transformation:
\begin{align}
\psi_{T^2/{\mathbb{Z}}_N^m,+}^{(j+\alpha_1,\alpha_2)}(z) = \psi_{T^2/{\mathbb{Z}}_N^m,+}^{(j+\alpha_1,\alpha_2)}(Z+z^{\rm{fp}}_I) = V_{z^{\rm{fp}}_I}(Z) \psi_{T^2/{\mathbb{Z}}_N^m,+}^{(j+\beta_1,\beta_2)}(Z),
\label{rightOBC1}
\end{align}
where $(\beta_1,\beta_2)$ are defined by
\begin{align}
(\beta_1,\beta_2)\equiv(\alpha_1+My^{\rm{fp}}_{2I},\alpha_2-My^{\rm{fp}}_{1I}) \quad({\rm{mod}}\,\,1).
\label{beta}
\end{align}
On the other hand, the left-hand side of Eq.\eqref{OBC1} can be written by $Z$ as
\begin{align}
\psi_{T^2/{\mathbb{Z}}_N^m,+}^{(j+\alpha_1,\alpha_2)}(\rho z)
&= \psi_{T^2/{\mathbb{Z}}_N^m,+}^{(j+\alpha_1,\alpha_2)}(\rho Z+\rho z^{\rm{fp}}_I) \notag \\
&= \psi_{T^2/{\mathbb{Z}}_N^m,+}^{(j+\alpha_1,\alpha_2)}(\rho Z+z^{\rm{fp}}_I-u-v\tau) \notag \\
&= U_2^{-v}(\rho Z+z^{\rm{fp}}_I-u) U_1^{-u}(\rho Z+z^{\rm{fp}}_I) V_{z^{\rm{fp}}_I}(\rho Z) \psi_{T^2/{\mathbb{Z}}_N^m,+}^{(j+\beta_1,\beta_2)}(\rho Z),
\label{leftOBC1}
\end{align}
where we use Eq.\eqref{FP}.
Thus, the mode functions $\psi_{T^2/{\mathbb{Z}}_N^m,+}^{(j+\beta_1,\beta_2)}(Z)$ transform under $\mathbb{Z}_N$ twist around $z^{\rm{fp}}_I$ as
\begin{align}
\psi_{T^2/{\mathbb{Z}}_N^m,+}^{(j+\beta_1,\beta_2)}(\rho Z)
&=\rho^{\chi_{+l}}
 \psi_{T^2/{\mathbb{Z}}_N^m,+}^{(j+\beta_1,\beta_2)}( Z),
 \label{WD}
\end{align}
with
\begin{align}
\chi_{+I}=N\{u\alpha_1+v\alpha_2+\tfrac{M}{2}(uv+u y^{\rm{fp}}_{2I} -v y^{\rm{fp}}_{1I})\}+m \quad ({\rm{mod}}\,\,N),
\label{WDN}
\end{align}
where we use the result,
\begin{align}
V_{z^{\rm{fp}}_I}^{-1}(\rho Z) V_{z^{\rm{fp}}_I}(Z) = e^{-\pi iM \frac{{\rm Im}((\bar{z}^{\rm{fp}}_I - \bar{\rho}\bar{z}^{\rm{fp}}_I) \rho Z)}{{\rm Im}\tau}} = U_1^{-u}(\rho Z) U_2^{-v}(\rho Z) e^{2\pi i (u\alpha_1+v\alpha_2)}.
\end{align}
Note that Eq.\eqref{WD} with $z^{\rm{fp}}_I = 0$ corresponds to Eq.\eqref{OBC1}.
Hence, we get the winding numbers $\chi_{+I}$ of the ${\mathbb{Z}}_N$ mode functions $\psi_{T^2/{\mathbb{Z}}_N^m,+}^{(j+\alpha_1,\alpha_2)}(z)$ around the fixed points $z^{\rm{fp}}_I$.

We are interested in the numbers of chiral zero modes on the $T^2/{\mathbb{Z}}_N$ orbifolds with magnetic flux. Although the AS index theorem on $T^2$ is known as \eqref{IndTorus}, the AS index theorem cannot, however, be applied to orbifolds directly because they have singular points. On the other hand, in the previous paper~\cite{Sakamoto:2020pev}, the following zero-mode counting formula on the $T^2/{\mathbb{Z}}_N$ orbifolds with magnetic flux has been obtained:
\begin{align}
n_{+}-n_{-}=\frac{M}{N}-\frac{V_+}{N}+1,
\label{Zeromodecounting}
\end{align}
where $V_{+}$ is the sum of the winding numbers at the fixed points of the $T^2/{\mathbb{Z}}_N$ orbifolds. 
It should be emphasized that the equality between the left-hand side and the right-hand side of Eq.\eqref{Zeromodecounting} has been verified in each case in Ref.~\cite{Sakamoto:2020pev}, but the formula \eqref{Zeromodecounting} has not been established as an index theorem. The first term on the right-hand side of Eq.\eqref{Zeromodecounting} can be understood as the contribution of the flux and the factor $1/N$ comes from the fact that the area of the $T^2/{\mathbb{Z}}_N$ orbifold is $1/N$ of that of the torus $T^2$. On the other hand, physical roles of the second and the third terms of Eq.\eqref{Zeromodecounting} are unclear, because they are not related to any flux on the orbifolds. In particular, it is curious why the factor $+1$ is needed on the right-hand side of the formula \eqref{Zeromodecounting}.

\begin{figure}[!t]
\centering
\includegraphics[width=0.6\textwidth]{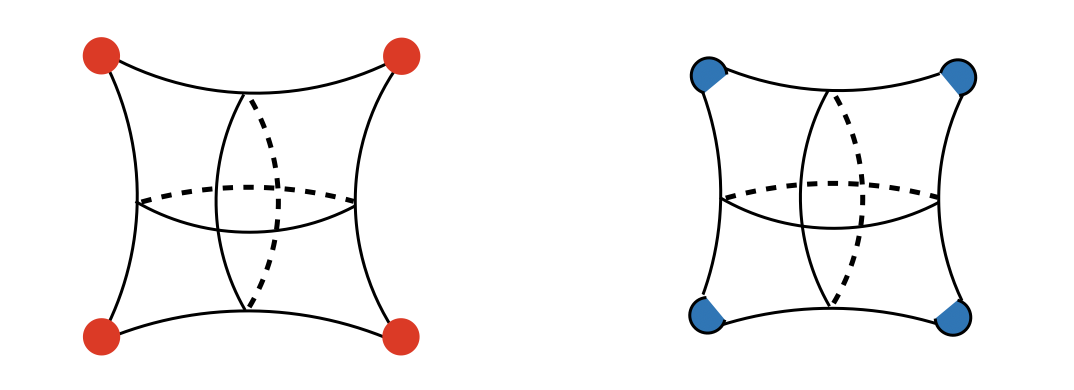}
\caption{The left figure shows $T^2/{\mathbb{Z}}_2$ orbifold and the red points represent the fixed points of $T^2/{\mathbb{Z}}_2$ orbifold. By cutting around the fixed points and embedding the part of $S^2$ as caps, we can construct the blow-up manifold as shown in the right figure.}
\label{blowup0}
\end{figure}

In order to apply the AS index theorem to the orbifold models, we consider removing the singular points from the orbifolds. To this end, we replace the $T^2/{\mathbb{Z}}_N$ orbifolds with smooth manifolds without singularities by cutting out the singularities of the magnetized $T^2/{\mathbb{Z}}_N$ orbifolds and attaching smooth manifolds (parts of $S^2$) to them, as shown in Figure \ref{blowup0}. The smooth manifolds without singularities are called blow-up manifolds of the $T^2/{\mathbb{Z}}_N$ orbifolds. Then, we can apply the AS index theorem to the blow-up manifolds directly.


\section{Blow-up manifold of magnetized $T^2/{\mathbb{Z}}_N$ orbifold}
In this section, to apply the AS index theorem to the orbifold models, we construct the blow-up manifolds of the magnetized $T^2/{\mathbb{Z}}_N$ orbifolds. We then need to connect wave functions on the orbifolds with those on parts of $S^2$ without losing the orbifold information.
In this analysis, it turns out that winding numbers of wave functions on the $T^2/{\mathbb{Z}}_N$ orbifolds are related to localized flux and localized curvature on the blow-up manifolds.
\color{black}
\subsection{Magnetized $S^2$}
\begin{figure}[!t]
\centering
\includegraphics[width=0.4\textwidth]{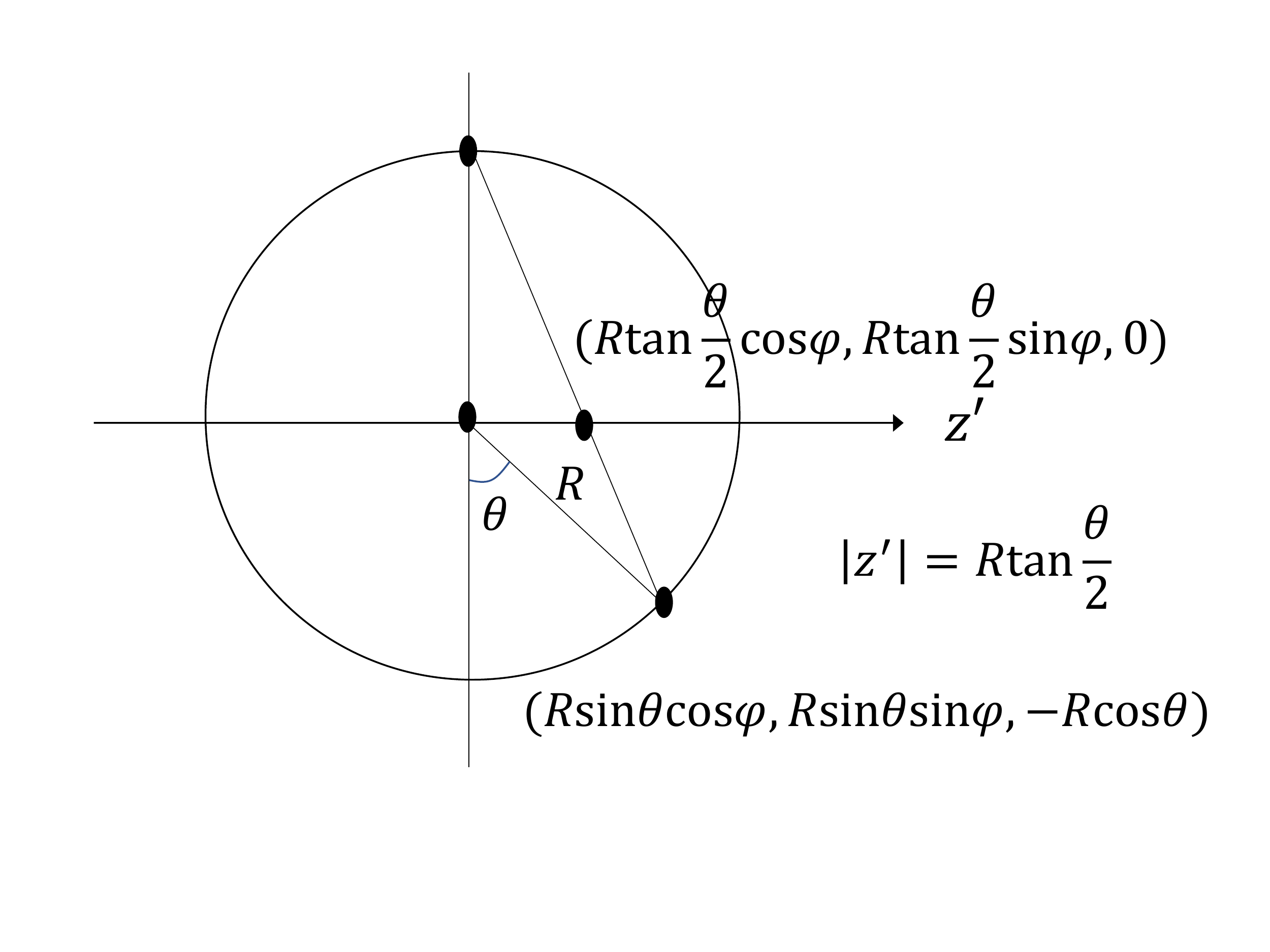}
\caption{The cross section of $S^2$ with the radius $R$ is shown. We project a point of $S^2$ with the 3d coordinate, $(R \sin \theta \cos \varphi, R \sin \theta \sin \varphi, -R \cos \theta)$, from the north pole of $S^2$, into the point on the complex plane passing through the center of $S^2$ whose 3d coordinate is $(R \tan \frac{\theta}{2} \cos \varphi, R \tan \frac{\theta}{2} \sin \varphi,0)$, where $(R, \theta, \varphi)$ are spherical coordinate parameters. We define the complex coordinate of the complex plane $\mathbb{CP}^1$, $z'$, such that $z' = |z'| e^{i\varphi} = R \tan \frac{\theta}{2} e^{i\varphi}$ at the point with the 3d coordinate $(R \tan \frac{\theta}{2} \cos \varphi, R \tan \frac{\theta}{2} \sin \varphi,0)$. Then, we denote the coordinate of a point on $S^2$ as the complex coordinate of the projected point on $\mathbb{CP}^1$, $z'$.}
\label{figureS2}
\end{figure}

In this subsection, we review zero mode functions on $S^2$ with magnetic flux~\cite{Conlon:2008qi}.
Let $z^{\prime}$ be the complex coordinate on $S^2\simeq {\mathbb{CP}}^1$ defined by projecting a point of $S^2$ into the complex plane passing through the center of $S^2$ from the north pole of $S^2$, as shown in Figure \ref{figureS2}.
The radius of $S^2$ is taken to be $R$. 

The magnetic flux on $S^2$ is quantized as 
\begin{align}
\frac{1}{2\pi}\int_{S^2} F^{\prime}=M^{\prime},
\end{align}
where $M^{\prime}$ is an integer. The field strength is 
\begin{align}
\frac{F^{\prime}}{2\pi}=\frac{i}{2\pi}\frac{R^2M^{\prime}}{(R^2+|z^{\prime}|^2)^2}dz^{\prime}\wedge d{\bar{z}}^{\prime}. \label{S2F}
\end{align}
The gauge potentials on $S^2$ are given by 
\begin{align}
A_{\bar{z}^{\prime}}=\frac{i}{2}\frac{M^{\prime}}{R^2+|z^{\prime}|^2}z^{\prime}, 
\qquad
A_{{z}^{\prime}}=-\frac{i}{2}\frac{M^{\prime}}{R^2+|z^{\prime}|^2}{\bar{z}}^{\prime}.
\label{gauge}
\end{align}

The mode functions on the magnetized $S^2$ obey the Dirac equations 
\begin{align}
\frac{R^2+|z^{\prime}|^2}{R}i\left(\partial_{\bar{z}^{\prime}} +i\frac{1}{2}\omega_{\bar{z}^{\prime}}-iA_{\bar{z}^{\prime}}\right)\psi_{S^2,+,n}(z^{\prime})=m_n\psi_{S^2,{-},n}(z^{\prime}), \label{DEQS21}\\
\frac{R^2+|z^{\prime}|^2}{R}i\left(\partial_{{z}^{\prime}} -i\frac{1}{2}\omega_{{z}^{\prime}}-iA_{{z}^{\prime}}\right)\psi_{S^2,-,n}(z^{\prime})=m_n\psi_{S^2,{+},n}(z^{\prime}) 
\label{DEQS2}
\end{align}
with
\begin{align}
\omega_{\bar{z}^{\prime}}=\frac{i}{2}\frac{2}{R^2+|z^{\prime}|^2}z^{\prime}, 
\qquad
\omega_{{z}^{\prime}}=-\frac{i}{2}\frac{2}{R^2+|z^{\prime}|^2}{\bar{z}}^{\prime}.
\label{spin}
\end{align}
Here, $\omega_{\bar{z}^{\prime}}$ and $\omega_{{z}^{\prime}}$ are the spin connections that come from the non-vanishing curvature on $S^2$:
\begin{align}
\frac{1}{2\pi}\int_{S^2}R^{\prime}=\chi(S^2)=2.
\end{align}
Here, $R^{\prime}$ is the curvature on $S^2$ and $\chi$ is the Euler characteristic.
Note that the spin connections \eqref{spin} can be obtained by replacing the flux $M^{\prime}$ in the gauge potentials \eqref{gauge} by the Euler characteristic $\chi(S^2)=2$.

The positive chirality zero mode solutions of Eq.\eqref{DEQS21} with $m_n=0$ are given by
\begin{align}
\psi_{S^2,+,0}(z^{\prime})=\frac{f_{+}(z^{\prime})}{(R^2+|z^{\prime}|^2)^{\frac{M^{\prime}-1}{2}}},
\label{S2zero}
\end{align}
where $f_{+}(z^{\prime})$ is a holomorphic function of $z^{\prime}$. These solutions are normalizable and well-defined on $S^2$ only if $M^{\prime}>0$ and $f_{+}(z^{\prime})$ is expressed as a $(M^{\prime}-1)$th order polynomial, which means that the number of the independent solutions is $M^{\prime}$.
On the other hand, normalizable and well-defined negative chirality zero modes on $S^2$ are obtained in a similar way only if $M^{\prime}<0$, and an anti-holomorphic function $f_{-}(\bar{z}^{\prime})$ is expressed as a $(|M^{\prime}|-1)$th order polynomial.

The above results are consistent with the AS index theorem on the magnetized $S^2$, i.e.
\begin{align}
n_{+}-n_{-}=\frac{1}{2\pi}\int_{S^2} F^{\prime}=M^{\prime}.
\label{IndS2}
\end{align}
The number of the chiral zero modes turns out to be given by the flux quantization number $M^{\prime}$, as it should be. It is important to emphasize that although the flux and the curvature exist in the magnetized $S^2$ model, only the flux contributes to the AS index theorem, as mentioned in the introduction.

\subsection{The relation between winding number, localized flux, and localized curvature at the fixed point}
To directly apply the AS index theorem, manifolds have to be smooth without singularities. Since the $T^2/{\mathbb{Z}}_N$ orbifolds have the singularities at the fixed points, we replace cones around the fixed points with parts of $S^2$ to remove the singularities (see Figure \ref{blowup0}). Then, it is important to construct the smooth blow-up manifolds without losing the orbifold information. In particular, it is crucial to preserve the information on winding numbers of wave function at the fixed points in the blow-up process. To realize it, we use a singular gauge transformation to connect wave functions on $T^2/{\mathbb{Z}}_N$ to those on $S^2$, as we will see later.
\begin{figure}[!t]
  \centering
\includegraphics[width=0.6\textwidth]{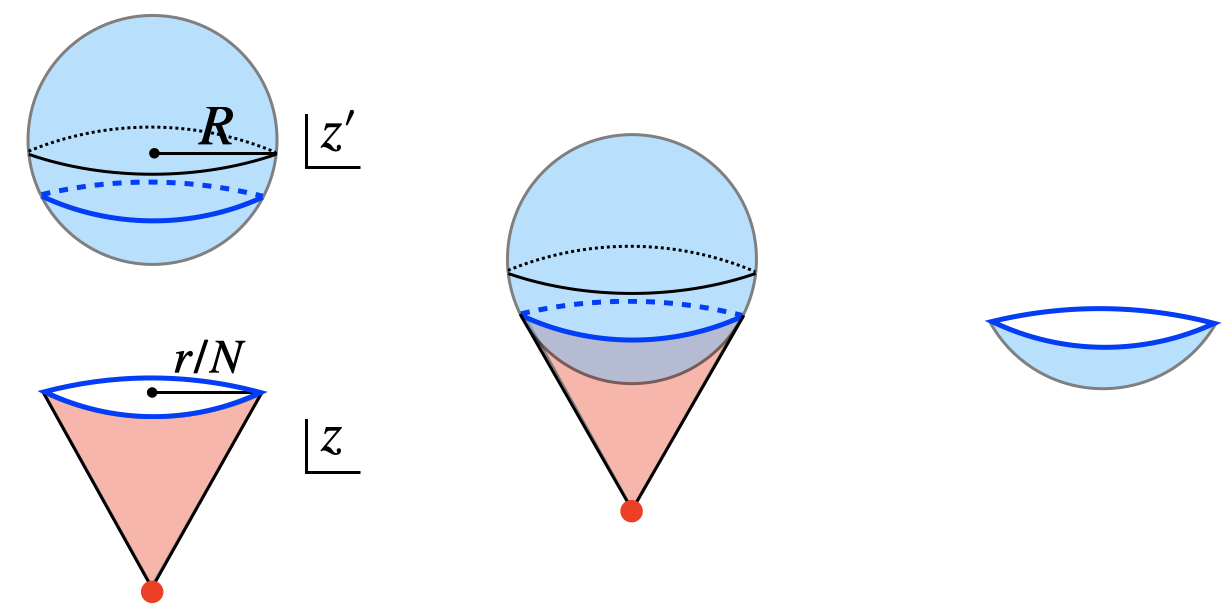}
    \caption{Cutting around the fixed point on the orbifold with a radius $r$ yields a cone (shown as a red cone in this figure) with a base of radius $r/N$. The magnetized $S^2$ (shown as a blue ball in this figure) is embedded in this  cone, aligning it with the connection line represented by the blue line. The $(N-1)/2N$-part of $S^2$ with the radius $R=r/\sqrt{N^2-1}$ is embedded by this operation. Here, $z$ and $z^{\prime}$ denote the coordinates of $T^2/{\mathbb{Z}}_N$ and $S^2$, respectively. The coordinates at the connection points are  $z=re^{i\varphi/N}$ and $z^{\prime}=re^{i\varphi}/(N+1)$, respectively.}
    \label{blowup123}
\end{figure}

First, we construct the blow-up manifolds by cutting around the singularities of the $T^2/{\mathbb{Z}}_N$ orbifolds and replacing them with parts of the magnetized $S^2$, as shown in Figures \ref{blowup0} and \ref{blowup123}. This replacement should be performed for each fixed point on the orbifolds. Details of the construction method of blow-up manifolds are discussed in~\cite{Kobayashi:2019fma} and specific relations of coordinates at the connection points are omitted here. It is necessary to smoothly connect the zero modes \eqref{Zeromode+1} and \eqref{S2zero} on the connection line. There is, however, an obstacle.  If zero modes on the $T^2/{\mathbb{Z}}_N$ orbifold have non-zero winding numbers, they cannot be connected to zero modes on $S^2$ because the boundary conditions  for $\psi_{T^2/{\mathbb{Z}}_N^m,+,0}^{(j+\alpha_1,\alpha_2)}(z)$ around the fixed point are different from those of  $\psi_{S^2,+,0}(z^{\prime})$. $\psi_{T^2/{\mathbb{Z}}_N^m,+}^{(j+\alpha_1,\alpha_2)}(z)$ obey the boundary condition \eqref{WD}. On the other hand, $\psi_{S^2,+,0}(z^{\prime})$ have no phase. To resolve the above problem, we use singular gauge transformations and remove non-zero winding numbers from wave functions on $T^2/{\mathbb{Z}}_N$, as will be discussed below.

We perform a singular gauge transformation $\psi_{T^2/{\mathbb{Z}}_N^m,+}^{(j+\alpha_1,\alpha_2)}(z)\,\to\,\tilde{\psi}_{T^2/{\mathbb{Z}}_N^m,+}^{(j+\alpha_1,\alpha_2)}(z)$ such that $\tilde{\psi}_{T^2/{\mathbb{Z}}_N^m,+}^{(j+\alpha_1,\alpha_2)}(z)$ has no winding number:
\begin{align}
\psi_{T^2/{\mathbb{Z}}_N^m,+}^{(j+\alpha_1,\alpha_2)}(\rho z)&=\rho^m \psi_{T^2/{\mathbb{Z}}_N^m,+}^{(j+\alpha_1,\alpha_2)}(z) \quad \rightarrow
\quad
\tilde{\psi}_{T^2/{\mathbb{Z}}_N^m,+}^{(j+\alpha_1,\alpha_2)}(\rho z)=
 \tilde{\psi}_{T^2/{\mathbb{Z}}_N^m,+}^{(j+\alpha_1,\alpha_2)}(z).
 \end{align}
Here, we have considered the case where the fixed point is $z^{\rm{fp}}_I=0$. Note that the following analysis can be applied even for the other fixed points by the following replacement:
\begin{align}
z&\to Z,  \\
(\alpha_1,\alpha_2) &\to (\beta_1,\beta_2),  \\
m &\to \chi_{+I}.
\end{align}
The singular gauge transformation is defined by
\begin{align}
&A \, \to \,\tilde{A}(z)\equiv A(z)+\delta A(z), \\
&\delta A(z)=\delta A_z dz +\delta A_{\bar{z}} d{\bar{z}}=i U_{\xi^F} d U_{\xi^F}^{-1}\simeq
-i\frac{\xi^F}{2}\frac{1}{z} dz+i\frac{\xi^F}{2}\frac{1}{\bar{z}} d{\bar{z}},
\label{localgauge}
\end{align}
with
\begin{align}
U_{\xi^F}(z)=\left(\frac{g_1(z)}{({g_1(z)})^{\ast}}\right)^{\frac{\xi^F}{2}} \simeq \left(\frac{g_1^{\prime}(0)z}{({g_1^{\prime}(0) z})^{\ast}}\right)^{\frac{\xi^F}{2}},
\label{UxiF}
\end{align}
where $g_1(z)$ denotes a specific holomorphic function with $\mathbb{Z}_N$ eigenvalue $1$.
The detailed form of $g_1(z)$ is discussed in \cite{Kobayashi}. The rightest-hand sides of Eqs.\eqref{localgauge} and \eqref{UxiF} are approximate expressions near $z=0$. Under the singular gauge transformation, the field strength is modified as
\begin{align}
&\frac{F}{2\pi}\,\,\to\,\, \frac{\tilde{F}}{2\pi}\equiv \frac{{F}}{2\pi}+\frac{\delta{F}}{2\pi}, \label{Ftilde} \\
&\frac{\delta F}{2\pi}=i\xi^F \delta(z)\delta(\bar{z})dz \wedge d{\bar{z}}. \label{deltaF}
\end{align}
We note that from Eq.\eqref{Ftilde} $\xi^F/N$ can be regarded as a localized flux at the fixed point $z=0$.

We further need to consider a singular gauge transformation for the spin connection in a way similar to the gauge potentials to remove winding numbers both of $\psi_{T^2/{\mathbb{Z}}_N^m,+}^{(j+\alpha_1,\alpha_2)}( z)$ and $\psi_{T^2/{\mathbb{Z}}_N^m,-}^{(j+\alpha_1,\alpha_2)}(z)$. It is defined by
\begin{align}
&\omega \,\to\,{\tilde{\omega}}=\omega+\delta \omega=\delta \omega, \quad (\omega = 0), \\
&\delta {\omega}=i U_{\xi^R} d U_{\xi^R}^{-1}\overset{z\simeq 0}{\simeq}
-i\frac{\xi^R}{{2}}\frac{1}{z} dz+i\frac{\xi^R}{{2}}\frac{1}{\bar{z}} d{\bar{z}},
\label{SGTcurvature}
\end{align}
with
\begin{align}
U_{\xi^R}(z)=\left(\frac{g_1(z)}{{(g_1(z))^{\ast}}}\right)^{\frac{\xi^R}{{2}}}\overset{z\simeq 0}{\simeq}\left(\frac{g_1^{\prime}(0)z}{({g_1^{\prime}(0) z})^{\ast}}\right)^{\frac{\xi^R}{{2}}}.
\label{UxiR}
\end{align}
We also note that $\xi^R/N$ can be regarded as a localized curvature at the fixed point. 
$\xi^R/N$ is given by the deficit angle as
\begin{align}
2\pi \frac{\xi^R}{N}=2\pi-\frac{2\pi}{N}=2\pi \frac{N-1}{N}, \label{xiR}
\end{align}
at the ${\mathbb{Z}}_N$ fixed point. 

From Eqs.\eqref{UxiF} and \eqref{UxiR}, the wave functions are transformed under the singular gauge transformation as
\begin{align}
{\psi}_{T^2/{\mathbb{Z}}_N^m,+}^{(j+\alpha_1,\alpha_2)}(z)\,\,\to\,\,\tilde{\psi}_{T^2/{\mathbb{Z}}_N^m,+}^{(j+\alpha_1,\alpha_2)}(z)&=
U_{\xi^F}(z)U_{\xi^R}^{-1/2}(z) {\psi}_{T^2/{\mathbb{Z}}_N^m,+}^{(j+\alpha_1,\alpha_2)}(z), 
\label{BCSGT1} \\
{\psi}_{T^2/{\mathbb{Z}}_N^m,-}^{(j+\alpha_1,\alpha_2)}(z)\,\,\to\,\,\tilde{\psi}_{T^2/{\mathbb{Z}}_N^m,-}^{(j+\alpha_1,\alpha_2)}(z)&=
U_{\xi^F}(z)U_{\xi^R}^{1/2}(z) {\psi}_{T^2/{\mathbb{Z}}_N^m,-}^{(j+\alpha_1,\alpha_2)}(z).
\label{BCSGT2}
\end{align}
Then, Eqs.\eqref{OBC1} and \eqref{OBC} are modified such as
\begin{align}
\tilde{\psi}_{T^2/{\mathbb{Z}}_N^m,+}^{(j+\alpha_1,\alpha_2)}(\rho z)&=
\rho^{\xi^F-\frac{\xi^R}{2}+m} \tilde{\psi}_{T^2/{\mathbb{Z}}_N^m,+}^{(j+\alpha_1,\alpha_2)}(z), 
\label{BCSGT11} \\
\tilde{\psi}_{T^2/{\mathbb{Z}}_N^m,-}^{(j+\alpha_1,\alpha_2)}(\rho z)&=
\rho^{\xi^F+\frac{\xi^R}{2}+m+1} \tilde{\psi}_{T^2/{\mathbb{Z}}_N^m,-}^{(j+\alpha_1,\alpha_2)}(z).
\label{BCSGT22}
\end{align}
Note that the contributions of the localized curvature $\xi^R$ act with opposite signs to the chirality positive and negative wave functions.

We arrive at the conditions to obtain wave functions with vanishing winding numbers as 
\begin{align}
\xi^F=\frac{N-1}{2}-m+\ell N \qquad {\rm{for}}\quad  ^{\forall} \ell \, \in \,{\mathbb{Z}},
\label{nowinding}
\end{align}
where we used $\xi^R = N-1$ for the ${\mathbb{Z}}_N$ fixed point.
It is interesting to point out that a new degree of freedom $\ell$ appears. It comes from ${\rm{mod}} \,N$ property of Eqs.\eqref{BCSGT11} and \eqref{BCSGT22}. 
For $z^{\rm{fp}}_I \neq 0$, the same argument can be applied by replacing $m$ with the winding number $\chi_{+I}$ at the fixed point $z^{\rm{fp}}_I$, yielding the following relationship:
 \begin{align}
\xi^F_I=\frac{N-1}{2}-{\chi_{+I}}+\ell_I N \qquad {\rm{for}}\quad  ^{\forall} \ell_I \, \in \,{\mathbb{Z}}.
\label{nowinding2}
\end{align}
Eq.\eqref{nowinding2} implies that the winding number $\chi_{+I}$ can be rewritten in terms of the localized flux $\xi^F_I$ and the localized curvature $\xi^R_I=N-1$. In other words, what we have done with the singular gauge transformations \eqref{BCSGT1} and \eqref{BCSGT2} is to replace the information of the winding number on the orbifolds with the localized flux and localized curvature at the fixed points of $T^2/{\mathbb{Z}}_N$. This operation is expected to connect wave functions on $T^2/{\mathbb{Z}}_N$ with those on $S^2$ without losing the orbifold information.
We will see in the next section that it allows us to reinterpret the index formula \eqref{Zeromodecounting}, which is one of our purposes in this paper.

\subsection{Flux condition}
Now, we can explore zero-mode wave functions on the blow-up manifolds of magnetized $T^2/{\mathbb{Z}}_N$.
Wave functions on the blow-up regions (parts of $S^2$ regions) are those on $S^2$ in Eq.\eqref{S2zero}, $\psi_{S^2,+,0}(z^{\prime})$, while those on the bulk region (remaining region of $T^2/{\mathbb{Z}}_N$ by  cutting out regions around fixed points) are those on $T^2/{\mathbb{Z}}_N$ in Eq.\eqref{BCSGT1}, $\tilde{\psi}_{T^2/{\mathbb{Z}}_N^m,+,0}^{(j+\alpha_1,\alpha_2)}(z)$, with the localized curvature \eqref{xiR} and the localized flux \eqref{nowinding}. 
In particular, the $\tilde{\psi}_{T^2/{\mathbb{Z}}_N^m,+}^{(j+\alpha_1,\alpha_2)}(z)$ near $z=0$ is {approximated} as
\begin{align}
\tilde{\psi}_{T^2/{\mathbb{Z}}_N^m,+,0}^{(j+\alpha_1,\alpha_2)}(z)
\simeq|z|^{m-\ell N}e^{-\frac{\pi M}{2{\rm{Im}}\tau}|z|^2 }|g_1^{\prime}(0)|^{m-\ell N}\tilde{h}_{+}^j(z),
\label{tildezero1}
\end{align}
where $\tilde{h}_{+}^j(z)$ denotes the holomorphic function.
Then, we should connect wave functions on bulk regions \eqref{tildezero1} and those on the blow-up regions \eqref{S2zero} smoothly at the junction points.
That is, 
$\tilde{\psi}_{T^2/{\mathbb{Z}}_N^m,+,0}^{(j+\alpha_1,\alpha_2)}(z)$ at $z=re^{i\varphi/N}$ and $\psi_{S^2,+,0}(z^{\prime})$ at $z^{\prime}=\frac{r}{N+1}e^{i\varphi}$ should satisfy the following junction conditions:
\begin{align}
\tilde{\psi}_{T^2/{\mathbb{Z}}_N^m,+,0}^{(j+\alpha_1,\alpha_2)}(z) \Big|_{z=re^{i\varphi/N}}
&=\psi_{S^2,+,0}(z^{\prime})\Big|_{z^{\prime}=\frac{r}{N+1}e^{i\varphi}}, \label{CC01}\\
\frac{1}{e^{-i\varphi/N}}\frac{d \tilde{\psi}_{T^2/{\mathbb{Z}}_N^m,+,0}^{(j+\alpha_1,\alpha_2)}(z)}{dz}\Big|_{z=re^{i\varphi/N}}&=
\frac{1}{\frac{N+1}{N}e^{-i\varphi}}\frac{d\psi_{S^2,+,0}(z^{\prime})}{dz^{\prime}}\Big|_{z^{\prime}=\frac{r}{N+1}e^{i\varphi}}.
\label{CC02}
\end{align}
The detailed analysis is discussed in Ref.~\cite{Kobayashi}. In particular,
from the non-holomorphic parts of Eqs.\eqref{CC01} and \eqref{CC02},  
we obtain the following relation:
\begin{align}
\frac{\pi r^2}{N{\rm{Im}}\tau}M+\frac{N-1}{2N}-\frac{m}{N}+\ell=\frac{N-1}{2N}M^{\prime}.
\label{CC}
\end{align}
For $z^{\rm{fp}}_I \neq 0$, by replacing $m$ with the winding number $\chi_{+I}$ we get
\begin{align}
\frac{\pi r^2_I}{N{\rm{Im}}\tau}M+\frac{N-1}{2N}-\frac{\chi_{+I}}{N}+\ell_I=\frac{N-1}{2N}M^{\prime}_I.
\label{CC2}
\end{align}

By using the relation \eqref{nowinding2} (or \eqref{nowinding}), Eq.\eqref{CC2} can be expressed as
\begin{align}
\frac{\pi r^2_I}{N{\rm{Im}}\tau}M + \frac{\xi^F_I}{N} = \frac{N-1}{2N}M^{\prime}_I.
\label{CC3}
\end{align}
Now, we can easily understand the physical meaning of this relationship. The left-hand side represents the flux, including the localized flux on the cutout area around the fixed point of $T^2/\mathbb{Z}_N$, while the right-hand side represents the flux on the embedded area of $S^2$. Thus, it means that the magnetic flux is not modified under the blow-up process.
This is important in deriving the AS index theorem, as we will see in the next section.
In the orbifold limit $r_I \to 0$, in particular, Eq.\eqref{CC3} is expressed as
\begin{align}
\frac{\xi^F_I}{N} = \frac{N-1}{2N}M^{\prime}_I \biggl|_{r_I=0},
\label{CC4}
\end{align}
which {shows} that the flux on the embedded area of $S^2$ (right-hand side of Eq.\eqref{CC4}) corresponds to the localized flux on the orbifold fixed point (left-hand side of Eq.\eqref{CC4}).

We notice that the curvature is not also modified under the blow-up process:
\begin{align}
\frac{\xi^R_I}{N} = \frac{N-1}{N} = \frac{N-1}{2N} \times 2,
\end{align}
where the left-hand side represents the curvature 
which corresponds to the deficit angle at the fixed point on
$T^2/\mathbb{Z}_N$, while the right-hand side represents the curvature in the embedded area of $S^2$.

\section{Index theorem on the blow-up manifold}
This section is the main section of this paper. Our purpose is to establish the AS index theorem on the $T^2/{\mathbb{Z}}_N$ orbifolds with magnetic flux background. Due to the existence of singularities on the orbifolds, the AS index theorem cannot be applied directly to the orbifold models. Our strategy is to replace the $T^2/{\mathbb{Z}}_N$ orbifolds with the blow-up manifolds without singularities and to apply the AS index theorem to them.

\subsection{Index theorem on the blow-up manifold}
The AS index theorem on the blow-up manifolds can be obtained as
\begin{align}
n_{+}-n_{-}&=\int_{{\rm{blow-up\, manifold}}} \frac{{F}}{2\pi} \label{Ind01} \\
&=\int_{{{T^2/{\mathbb{Z}}_N}\,\rm{bulk}}}\frac{F}{2\pi}+
\sum_{I}\int_{\frac{N-1}{2N}\times S^2}\frac{{F}^{\prime}}{2\pi} \label{Ind02} \\
&=\left( \frac{M}{N} -\sum_{I} \frac{\pi r_I^2}{N{\rm{Im}}\tau}M \right) +\sum_{I}\frac{N-1}{2N} M^{\prime}_I(r_I) \label{Ind03} \\
&=\left( \frac{M}{N} -\sum_{I} \frac{\pi r_I^2}{N{\rm{Im}}\tau}M \right) + \sum_{I} \left( \frac{\pi r^2_I}{N{\rm{Im}}\tau}M + \frac{\xi^F_I}{N} \right) \label{Ind04} \\
&= \frac{M}{N} + \sum_{I} \frac{\xi^F_I}{N}. \label{Ind05}
\end{align}

There are several comments for the above equations.
For Eq.\eqref{Ind01}, we emphasize that the index $n_{+}-n_{-}$ on the blow-up manifolds does not depend on the curvature but only on the flux.
It comes from the fact that the AS index theorem on a two-dimensional compact manifold has only the contribution of the flux on the manifold, in general.
For the first term of Eq.\eqref{Ind02} (and Eq.\eqref{Ind03}), the $T^2/{\mathbb{Z}}_N$ bulk refers to the region of the $T^2/{\mathbb{Z}}_N$ orbifold from which the areas near the fixed points are removed. 
For the second term of Eq.\eqref{Ind02} (and Eq.\eqref{Ind03}), it represents each amount of the magnetic flux on the embedded area of $S^2$ replacing the fixed point. The sum over $I$ is taken for the fixed points of the $T^2/{\mathbb{Z}}_N$ orbifolds. 
For Eq.\eqref{Ind04}, we used the relation \eqref{CC3}.

For the final result \eqref{Ind05}, it should be
emphasized that the AS index theorem on the blow-up manifolds does not depend on the blow-up radius $r_I$, as it should be.
In other words, the result of the AS index theorem holds even in the orbifold limit $r_I \to 0$:
\begin{align}
n_{+}-n_{-}=\int_{T^2/{\mathbb{Z}}_N} \frac{\tilde{F}}{2\pi}=\frac{M}{N}+\sum_{I}\frac{\xi^F_{I}}{N}.
\label{Ind2}
\end{align}
Here, $\tilde{F}$ is defined in Eq.\eqref{Ftilde} and this term comes from the limit of the right-hand side of Eq.\eqref{Ind02} as follows: in the $r_I \to 0$ ($R \to 0$) limit, the second term of Eq.\eqref{Ind02} with the field strength \eqref{S2F} can be expressed as
\begin{align}
\int_{\frac{N-1}{2N}\times S^2} i M^{\prime}_I \delta(z^{\prime})\delta(\bar{z}^{\prime}) dz^{\prime} \wedge d{\bar{z}}^{\prime},
\end{align}
(see Appendix A in Ref.~\cite{Bershadsky:1993cx}),
and it corresponds to Eq.\eqref{deltaF} by considering Eq.\eqref{CC4}. Thus,
the first term of the rightest-hand side of Eq.\eqref{Ind2} represents the contribution of the homogeneous magnetic flux on the $T^2/{\mathbb{Z}}_N$ orbifolds, which comes from the first term of \eqref{Ftilde}, while the second term represents the sum of localized fluxes at each fixed point, which comes from the second term of \eqref{Ftilde}. 
Therefore, Eq.\eqref{Ind2} becomes the AS index theorem on the $T^2/{\mathbb{Z}}_N$ orbifold, and the index can be determined by only the contribution of the flux.

From Eq.\eqref{nowinding2}, 
the localized flux $\xi^F_I$ is decided by the localized curvature $\xi^R_I=N-1$ and the winding number $\chi_{+I}$ at the fixed points. The winding numbers at the fixed points are investigated in~\cite{Sakamoto:2020pev}, and we can derive the values of the localized flux.
We can verify that the number of chiral zero modes, which are computed by the zero-mode counting formula in Ref.~\cite{Sakamoto:2020pev}, are completely consistent with the relation \eqref{Ind2}. The results are summarized in Table \ref{T1}$-$\ref{Z6tab} of the appendix. Although it was not clear whether the zero mode counting formula was the AS index theorem, the present results using the blow-up manifolds indicate that it is indeed the case.

\subsection{Reinterpretation of the zero-mode counting formula}
We can now reinterpret the zero-mode counting formula \eqref{Zeromodecounting}. 
Using the relation \eqref{nowinding2}, the AS index theorem \eqref{Ind05} can be rewritten in terms of the winding numbers $\chi_{+I}$ as
\begin{align}
n_{+}-n_{-}=\frac{M}{N}+\sum_{I}\left(\frac{-\chi_{+I}}{N}+\frac{N-1}{2N}+\ell_I \right)
=\frac{M-V_{+}}{N}+1+\sum_{I} \ell_I.
\label{Ind3}
\end{align}
Here, we have used the relation
\begin{align}
\sum_{I}\frac{N-1}{2N}=\frac{1}{2}\sum_{I}\frac{\xi^R_I}{N} =1,
\end{align}
at the last equality. It can be verified as follows: 
\begin{align}
&T^2/{\mathbb{Z}}_2 \,: \,\,\sum_{I}\frac{\xi^R_I}{4}=4\times \frac{1}{4}=1 \qquad \left(
z^{\rm{fp}}_I=
0, \frac{1}{2},\frac{\tau}{2},\frac{1+\tau}{2}\right), \\
&T^2/{\mathbb{Z}}_3 \,:\,\, \sum_{I}\frac{\xi^R_I}{6}=3\times \frac{2}{6}=1 \qquad \left(
z^{\rm{fp}}_I=
0, \frac{2+\tau}{3},\frac{1+2\tau}{3}\right), \\
&T^2/{\mathbb{Z}}_4 \,: \,\,\sum_{I_{{\mathbb{Z}}_4}}\frac{\xi^R_{I_{\mathbb{Z}_4}}}{8}+\frac{1}{2}\sum_{I_{{\mathbb{Z}}_2}}\frac{\xi^R_{I_{\mathbb{Z}_2}}}{4}=2\times \frac{3}{8}+\frac{1}{2}\times 2\times \frac{1}{4}=1 \, \left(
z^{\rm{fp}}_{I_{{\mathbb{Z}}_4}}=
0,\frac{1+\tau}{2},\,z^{\rm{fp}}_{I_{{\mathbb{Z}}_2}}=\frac{1}{2},\frac{\tau}{2}
\right), \\
&T^2/{\mathbb{Z}}_6 \,:\,\, \sum_{I_{{\mathbb{Z}}_6}}\frac{\xi^R_{I_{\mathbb{Z}_6}}}{12}+\frac{1}{2}\sum_{I_{{\mathbb{Z}}_3}}\frac{\xi^R_{I_{\mathbb{Z}_3}}}{6}+
\frac{1}{3}\sum_{I_{{\mathbb{Z}}_2}}\frac{\xi^R_{I_{\mathbb{Z}_2}}}{4}
=\frac{5}{12}+\frac{2}{6}+\frac{1}{4}=1 
\notag \\&\qquad\qquad \left(
z^{\rm{fp}}_{I_{{\mathbb{Z}}_6}}=
0,\quad z^{\rm{fp}}_{I_{{\mathbb{Z}}_3}}=\frac{1+\tau}{3},\frac{2+2\tau}{3},\quad z^{\rm{fp}}_{I_{{\mathbb{Z}}_2}}=\frac{1}{2},\frac{\tau}{2},\frac{1+\tau}{2}
\right).
\end{align}
Note that ${\mathbb{Z}}_4$ and ${\mathbb{Z}}_6$ have subgroups and must include the contributions of their fixed points.

Thus, the zero-mode counting formula \eqref{Zeromodecounting} can be derived from Eq.\eqref{Ind3} by taking $\ell_I=0$. The physical meaning of $+1$ in Eq.\eqref{Zeromodecounting}, which had been a mystery, is now clear. The factor $+1$ is the contribution of the sum of the localized curvatures at fixed points. When we try to write the index theorem with the winding numbers, $+1$ is needed to remove the contribution of the localized curvature from them, since the winding numbers include the contributions of both localized flux and the localized curvature (see Eq.\eqref{nowinding2}). This analysis reveals that the zero-mode counting formula includes only the contribution of the flux. 

An interesting observation in our analysis is the existence of a new degree of freedom $\ell_I$. The AS index theorem says that additional zero modes can appear.  This will be discussed in detail in~\cite{Kobayashi}, so we will not go into it here.


\section{Conclusion}
In this paper, we have considered the blow-up manifolds of the $T^2/{\mathbb{Z}}_N\,\,(N=2,3,4,6)$ orbifolds with magnetic flux background to establish the AS index theorem {on} the orbifolds. In our previous paper~\cite{Sakamoto:2020pev}, we have got the zero-mode counting formula which gives the numbers of the chiral zero modes {on} $T^2/{\mathbb{Z}}_N$ orbifolds. It is, however, unclear whether the formula can be regarded as the index theorem, because the equality between the left-hand side and the right-hand side of Eq.\eqref{Zeromodecounting} was merely verified in Ref.~\cite{Sakamoto:2020pev}. Furthermore, it is not obvious why the sum of the winding numbers $V_{+}$ appears and what is the physical meaning of the factor $+1$ in the formula \eqref{Zeromodecounting}.

To confirm the zero-mode counting formula \eqref{Zeromodecounting} as the index theorem and also to reveal the physical and geometrical meanings of the right-hand side of the formula \eqref{Zeromodecounting}, we have constructed the blow-up manifolds without singularities from the $T^2/{\mathbb{Z}}_N$ orbifolds by cutting out around the singularities of the $T^2/{\mathbb{Z}}_N$ orbifolds and attaching smooth manifolds (parts of $S^2$) to them.

In Section 3, in the process of construction of the magnetized smooth manifolds (blow-up manifolds), we obtained two important conditions, \eqref{nowinding2} and \eqref{CC4}.
The first condition comes from the modification of boundary conditions of wave functions on the $T^2/{\mathbb{Z}}_N$ orbifolds by {the} appropriate singular gauge transformation, which is needed to connect wave functions on $T^2/{\mathbb{Z}}_N$ and those on $S^2$. This condition means that the contributions of the winding numbers can be written by those of the localized flux and the localized curvature at each fixed point.
The second condition comes from the junction conditions of wave functions on $T^2/{\mathbb{Z}}_N$ and those on $S^2$. This condition means that the magnetic flux (as well as the curvature) is not modified under the blow-up process.
This result becomes important for deriving the AS index theorem on the $T^2/{\mathbb{Z}}_N$ orbifolds.

In Section 4, we have applied the AS index theorem to the blow-up manifolds of the $T^2/{\mathbb{Z}}_N$ orbifolds, and the numbers of chiral zero modes are given only by the magnetic flux on the blow-up manifolds.
Since the total flux is not modified under the blow-up process, the result is unchanged even in the orbifold limit $r_I \to 0$, and 
the AS index theorem on $T^2/{\mathbb{Z}}_N$ orbifolds with magnetic flux background is expressed by Eq.\eqref{Ind2}.
It shows that the index is decided by the contribution of the homogeneous magnetic flux $M$ and the localized fluxes $\xi^F_I$ at the fixed points. We have verified that {the number of chiral zero modes obtained by} the zero-mode counting formula \eqref{Zeromodecounting} in~\cite{Sakamoto:2020pev} is completely consistent with Eq.\eqref{Ind2}. 
The zero-mode counting formula can be reinterpreted from the viewpoint of the blow-up manifolds. The factor $+1$ in the formula \eqref{Zeromodecounting} is found to be the contribution of the localized curvature at the fixed points and is needed to remove the contribution of the localized curvature from the winding numbers because the winding numbers include the contributions of both the localized flux and the localized curvature. (Remember that the AS index theorem in two dimensions needs only the information of fluxes.)
\color{black}
Interestingly,
a new degree of freedom $\ell$ in Eq.\eqref{Ind3}, which emerges from the indeterminacy of mod $N$, suggests that there are new {$|\ell|$} number of chiral zero modes.
The new chiral zero modes will be discussed in detail in Ref.~\cite{Kobayashi}.

It will be important to extend our analysis for the AS index theorem on higher dimensional toroidal orbifolds such as $T^4/\mathbb{Z}_N$ and $T^6/\mathbb{Z}_N$.\footnote{See for the higher dimensional orbifold models with bulk magnetic fluxes \cite{Abe:2014nla,Kikuchi:2022lfv} as well as localized fluxes \cite{GrootNibbelink:2007lua,Leung:2019oln}.} We would study them elsewhere.

\section*{Acknowledgment}
This work was supported by JSPS KAKENHI Grants No. JP20K14477 (H. O.), JP 18K03649 (M. S.), JP 21J20739 (M. T.) and JP 20J20388 (H. U.), and the Education and Research
Program for Mathematical and Data Science from the Kyushu University (H. O.). 
Y.T. is supported in part by Scuola Normale, by INFN (IS GSS-Pi) and by the MIUR-PRIN contract 2017CC72MK\_003.

\appendix
\section{Localized flux and index}
We compared the value of the index obtained from Eq.\eqref{Ind2} with the result obtained from the zero-mode counting formula \cite{Sakamoto:2020pev} and confirmed that these are consistent in all cases. The results are summarized in Tables \ref{T1}--\ref{Z6tab}.

\begin{table}[!ht]
\centering
{\tabcolsep = 4mm
\renewcommand{\arraystretch}{1.2}
\scalebox{0.85}{
\begin{tabular}{ccc|c:c:c:c|c|c} \hline
flux & parity & twist & \multicolumn{4}{c|}{localized flux}  &index & \eqref{Zeromodecounting}\\
$M$ & $\eta$ & $(\alpha_1, \alpha_2)$ & $\frac{\xi^F_{1}}{N}$ & $\frac{\xi^F_{2}}{N}$ & $\frac{\xi^F_{3}}{N}$ & $\frac{\xi^F_{4}}{N}$ &  $\frac{M}{N}+\sum_{I=1}^4 \frac{\xi^F_I}{N}$ &$\frac{M-V_{+}}{N}+1$\\ \hline
$2m+1$ & $+1$ & $(0,0)$ & $1/4$ & $1/4$ & $1/4$ & $-1/4$& $(M+1)/2$&$(M+1)/2$\\
&& $(\tfrac12, 0)$ & $1/4$ & $-1/4$ & $1/4$ & $1/4$& $(M+1)/2$& $(M+1)/2$ \\
&& $(0, \tfrac12)$ & $1/4$ & $1/4$ & $-1/4$ & $1/4$& $(M+1)/2$& $(M+1)/2$ \\
&& $(\tfrac12, \tfrac12)$ & $1/4$ & $-1/4$ & $-1/4$ & $-1/4$& $(M-1)/2$  & $(M-1)/2$ \\ \cline{2-9}
& $-1$ & $(0,0)$ & $-1/4$ & $-1/4$ & $-1/4$ & $1/4$& $(M-1)/2$& $(M-1)/2$ \\
&& $(\tfrac12,0)$ & $-1/4$ & $1/4$ & $-1/4$ & $-1/4$& $(M-1)/2$ & $(M-1)/2$ \\
&& $(0, \tfrac12)$ & $-1/4$ & $-1/4$ & $1/4$ & $-1/4$& $(M-1)/2$& $(M-1)/2$ \\
&& $(\tfrac12, \tfrac12)$ & $-1/4$ & $1/4$ & $1/4$ & $1/4$& $(M+1)/2$& $(M+1)/2$ \\ \hline
$2m+2$ & $+1$ & $(0,0)$ & $1/4$ & $1/4$ & $1/4$ & $1/4$& $M/2+1$ & $M/2+1$ \\
&& $(\tfrac12, 0)$ & $1/4$ & $-1/4$ & $1/4$ & $-1/4$& $M/2$& $M/2$ \\
&& $(0, \tfrac12)$ & $1/4$ & $1/4$ & $-1/4$ & $-1/4$& $M/2$& $M/2$ \\
&& $(\tfrac12, \tfrac12)$ & $1/4$ & $-1/4$ & $-1/4$ & $1/4$& $M/2$& $M/2$ \\ \cline{2-9}
& $-1$ & $(0,0)$ & $-1/4$ & $-1/4$ & $-1/4$ & $-1/4$& $M/2-1$& $M/2-1$ \\
&& $(\tfrac12, 0)$& $-1/4$ & $1/4$ & $-1/4$ & $1/4$& $M/2$& $M/2$ \\
&& $(0, \tfrac12)$ & $-1/4$ & $-1/4$ & $1/4$ & $1/4$& $M/2$& $M/2$ \\
&& $(\tfrac12, \tfrac12)$ & $-1/4$ & $1/4$ & $1/4$ & $-1/4$& $M/2$ & $M/2$ \\ \hline
\end{tabular}
}
}
\caption{The values of localized fluxes at fixed points and index ($l=0$) on $T^2/{\mathbb{Z}}_2$.
}
\label{T1}
\end{table} 

\begin{table}[!ht]
	\centering
	{\tabcolsep = 3mm
		\renewcommand{\arraystretch}{1.2}
		\scalebox{1.00}{
			\begin{tabular}{ccc|c:c:c|c|c} \hline
			 flux& parity& twist & \multicolumn{3}{c|}{localized flux } & index& \eqref{Zeromodecounting}\\
				$M$ & $\eta$ & $\alpha$ & $\frac{\xi^F_{1}}{N}$ & $\frac{\xi^F_{2}}{N}$ & $\frac{\xi^F_{3}}{N}$ & $\frac{M}{N}+\sum_{I=1}^3 \frac{\xi^F_I}{N}$ &
				$\frac{M-V_{+}}{N}+1$\\ \hline
				$6m+1$ & $1$ & $1/6$ & $1/3$ & $0$ & $1/3$ & $(M+2)/3$ & $(M+2)/3$ \\
				&& $1/2$ & $1/3$ & $-1/3$ & $-1/3$ & $(M-1)/3$  & $(M-1)/3$ \\
				&& $5/6$ & $1/3$ & $1/3$ & $0$ & $(M+2)/3$ & $(M+2)/3$  \\ \cline{2-8}
				& $\omega$ & $1/6$ & $0$ & $-1/3$ & $0$ & $(M-1)/3$ & $(M-1)/3$  \\
				&& $1/2$ & $0$ & $1/3$ & $1/3$ & $(M+2)/3$ & $(M+2)/3$  \\
				&& $5/6$ & $0$ & $0$ & $-1/3$ & $(M-1)/3$ & $(M-1)/3$  \\ \cline{2-8}
				& $\omega^2$ & $1/6$ & $-1/3$ & $1/3$ & $-1/3$ & $(M-1)/3$ & $(M-1)/3$  \\
				&& $1/2$ & $-1/3$ & $0$ & $0$ & $(M-1)/3$ & $(M-1)/3$  \\
				&& $5/6$ & $-1/3$ & $-1/3$ & $1/3$ & $(M-1)/3$ & $(M-1)/3$  \\ \hline
				$6m+2$ & $1$ & $0$ & $1/3$ & $0$ & $0$ & $(M+1)/3$ & $(M+1)/3$  \\
				&& $1/3$ & $1/3$ & $-1/3$ & $1/3$ & $(M+1)/3$ & $(M+1)/3$  \\
				&& $2/3$ & $1/3$ & $1/3$ & $-1/3$ & $(M+1)/3$ & $(M+1)/3$  \\ \cline{2-8}
				& $\omega$ & $0$ & $0$ & $-1/3$ & $-1/3$ & $(M-2)/3$ & $(M-2)/3$  \\
				&& $1/3$ & $0$ & $1/3$ & $0$ & $(M+1)/3$ & $(M+1)/3$  \\
				&& $2/3$ & $0$ & $0$ & $1/3$ & $(M+1)/3$ & $(M+1)/3$  \\ \cline{2-8}
				& $\omega^2$ & $0$ & $-1/3$ & $1/3$ & $1/3$ & $(M+1)/3$ & $(M+1)/3$  \\
				&& $1/3$ & $-1/3$ & $0$ & $-1/3$ & $(M-2)/3$ & $(M-2)/3$  \\
				&& $2/3$ & $-1/3$ & $-1/3$ & $0$ & $(M-2)/3$ & $(M-2)/3$  \\ \hline
				$6m+3$ & $1$ & $1/6$ & $1/3$ & $-1/3$ & $0$ & $M/3$ & $M/3$  \\
				&& $1/2$ & $1/3$ & $1/3$ & $1/3$ & $M/3+1$ & $M/3+1$  \\
				&& $5/6$ & $1/3$ & $0$ & $-1/3$ & $M/3$ & $M/3$  \\ \cline{2-8}
				& $\omega$ & $1/6$ & $0$ & $1/3$ & $-1/3$ & $M/3$ & $M/3$  \\
				&& $1/2$ & $0$ & $0$ & $0$ & $M/3$ & $M/3$  \\
				&& $5/6$ & $0$ & $-1/3$ & $1/3$ & $M/3$ & $M/3$  \\ \cline{2-8}
				& $\omega^2$ & $1/6$ & $-1/3$ & $0$ & $1/3$ & $M/3$ & $M/3$  \\
				&& $1/2$ & $-1/3$ & $-1/3$ & $-1/3$ & $M/3-1$ & $M/3-1$  \\
				&& $5/6$ & $-1/3$ & $1/3$ & $0$ & $M/3$ & $M/3$  \\ \hline
			\end{tabular}
		}
	}
	\caption{The values of localized fluxes at fixed points and index ($l=0$) on $T^2/{\mathbb{Z}}_3$.
		}
	\label{Z3tab1}
\end{table}
\begin{table}[!ht]
	\centering
	{\tabcolsep = 3mm
		\renewcommand{\arraystretch}{1.2}
		\scalebox{1.00}{
			\begin{tabular}{ccc|c:c:c|c|c} \hline
			 flux& parity& twist & \multicolumn{3}{c|}{localized flux } & index& \eqref{Zeromodecounting}\\				$M$ & $\eta$ & $\alpha$ & $\frac{\xi^F_{1}}{N}$ & $\frac{\xi^F_{2}}{N}$ & $\frac{\xi^F_{3}}{N}$ & $\frac{M}{N}+\sum_{I=1}^3 \frac{\xi^F_I}{N}$ &
				$\frac{M-V_{+}}{N}+1$\\ \hline
				$6m+4$ & $1$ & $0$ & $1/3$ & $-1/3$ & $-1/3$ & $(M-1)/3$ & $(M-1)/3$ \\
				&& $1/3$ & $1/3$ & $1/3$ & $0$ & $(M+2)/3$  & $(M+2)/3$ \\
				&& $2/3$ & $1/3$ & $0$ & $1/3$ & $(M+2)/3$ & $(M+2)/3$  \\ \cline{2-8}
				& $\omega$ & $0$ & $0$ & $1/3$ & $1/3$ & $(M+2)/3$ & $(M+2)/3$  \\
				&& $1/3$ & $0$ & $0$ & $-1/3$ & $(M-1)/3$ & $(M-1)/3$  \\
				&& $2/3$ & $0$ & $-1/3$ & $0$ & $(M-1)/3$ & $(M-1)/3$  \\ \cline{2-8}
				& $\omega^2$ & $0$ & $-1/3$ & $0$ & $0$ & $(M-1)/3$ & $(M-1)/3$  \\
				&& $1/3$ & $-1/3$ & $-1/3$ & $1/3$ & $(M-1)/3$ & $(M-1)/3$  \\
				&& $2/3$ & $-1/3$ & $1/3$ & $-1/3$ & $(M-1)/3$ & $(M-1)/3$  \\ \hline
				$6m+5$ & $1$ & $1/6$ & $1/3$ & $1/3$ & $-1/3$ & $(M+1)/3$ & $(M+1)/3$  \\
				&& $1/2$ & $1/3$ & $0$ & $0$ & $(M+1)/3$ & $(M+1)/3$  \\
				&& $5/6$ & $1/3$ & $-1/3$ & $1/3$ & $(M+1)/3$ & $(M+1)/3$  \\ \cline{2-8}
				& $\omega$ & $1/6$ & $0$ & $0$ & $1/3$ & $(M+1)/3$ & $(M+1)/3$  \\
				&& $1/2$ & $0$ & $-1/3$ & $-1/3$ & $(M-2)/3$ & $(M-2)/3$  \\
				&& $5/6$ & $0$ & $1/3$ & $0$ & $(M+1)/3$ & $(M+1)/3$  \\ \cline{2-8}
				& $\omega^2$ & $1/6$ & $-1/3$ & $-1/3$ & $0$ & $(M-2)/3$ & $(M-2)/3$  \\
				&& $1/2$ & $-1/3$ & $1/3$ & $1/3$ & $(M+1)/3$ & $(M+1)/3$  \\
				&& $5/6$ & $-1/3$ & $0$ & $-1/3$ & $(M-2)/3$ & $(M-2)/3$  \\ \hline
				$6m+6$ & $1$ & $0$ & $1/3$ & $1/3$ & $1/3$ & $M/3+1$ & $M/3+1$  \\
				&& $1/3$ & $1/3$ & $0$ & $-1/3$ & $M/3$ & $M/3$  \\
				&& $2/3$ & $1/3$ & $-1/3$ & $0$ & $M/3$ & $M/3$  \\ \cline{2-8}
				& $\omega$ & $0$ & $0$ & $0$ & $0$ & $M/3$ & $M/3$  \\
				&& $1/3$ & $0$ & $-1/3$ & $1/3$ & $M/3$ & $M/3$  \\
				&& $2/3$ & $0$ & $1/3$ & $-1/3$ & $M/3$ & $M/3$  \\ \cline{2-8}
				& $\omega^2$ & $0$ & $-1/3$ & $-1/3$ & $-1/3$ & $M/3-1$ & $M/3-1$  \\
				&& $1/3$ & $-1/3$ & $1/3$ & $0$ & $M/3$ & $M/3$  \\
				&& $2/3$ & $-1/3$ & $0$ & $1/3$ & $M/3$ & $M/3$  \\ \hline
			\end{tabular}
		}
	}
	\caption{The values of localized fluxes at fixed points and index ($l=0$) on $T^2/{\mathbb{Z}}_3$.
		}
	\label{Z3tab2}
\end{table}

\begin{table}[!ht]
	\centering
	{\tabcolsep = 3mm
		\renewcommand{\arraystretch}{1.2}
		\scalebox{1.00}{
			\begin{tabular}{ccc|c:c:c:c|c|c} \hline
			 flux& parity& twist & \multicolumn{4}{c|}{localized flux } & index& \eqref{Zeromodecounting}\\
				$M$ & $\eta$ & $\alpha$ & $\frac{\xi^F_{1}}{N}$& $\frac{\xi^F_{2}}{N}$ & $\frac{\xi^F_{3}}{N}$ & $\frac{\xi^F_{4}}{N}$ & $\frac{M}{N}+\sum_{I=1}^4 \frac{\xi^F_I}{N}$ &
				$\frac{M-V_{+}}{N}+1$\\ \hline

				$4m+1$ & $1$ & $0$ & $3/8$ & $1/8$ &$1/8$ & $1/8$ & $(M+3)/4$ & $(M+3)/4$ \\
				& & $1/2$ & $3/8$ & $-3/8$ & $-1/8$ & $-1/8$ & $(M-1)/4$  & $(M-1)/4$  \\ \cline{2-9}
				& $i$ & $0$ & $1/8$ & $-1/8$ &$-1/8$ & $-1/8$ & $(M-1)/4$ & $(M-1)/4$ \\
				& & $1/2$ & $1/8$ & $3/8$ & $1/8$ & $1/8$ & $(M+3)/4$  & $(M+3)/4$  \\ \cline{2-9}
				 & $-1$ & $0$ & $-1/8$ & $-3/8$ &$1/8$ & $1/8$ & $(M-1)/4$ & $(M-1)/4$ \\
				& & $1/2$ & $-1/8$ & $1/8$ & $-1/8$ & $-1/8$ & $(M-1)/4$  & $(M-1)/4$  \\ \cline{2-9}
				& $-i$ & $0$ & $-3/8$ & $3/8$ &$-1/8$ & $-1/8$ & $(M-1)/4$ & $(M-1)/4$ \\
				& & $1/2$ & $-3/8$ & $-1/8$ & $1/8$ & $1/8$ & $(M-1)/4$  & $(M-1)/4$  \\ \hline	
				$4m+2$ & $1$ & $0$ & $3/8$ & $-1/8$ &$1/8$ & $1/8$ & $(M+2)/4$ & $(M+2)/4$ \\
				& & $1/2$ & $3/8$ & $3/8$ & $-1/8$ & $-1/8$ & $(M+2)/4$  & $(M+2)/4$  \\ \cline{2-9}
				& $i$ & $0$ & $1/8$ & $-3/8$ &$-1/8$ & $-1/8$ & $(M-2)/4$ & $(M-2)/4$ \\
				& & $1/2$ & $1/8$ & $1/8$ & $1/8$ & $1/8$ & $(M+2)/4$  & $(M+2)/4$  \\ \cline{2-9}
				 & $-1$ & $0$ & $-1/8$ & $3/8$ &$1/8$ & $1/8$ & $(M+2)/4$ & $(M+2)/4$ \\
				& & $1/2$ & $-1/8$ & $-1/8$ & $-1/8$ & $-1/8$ & $(M-2)/4$  & $(M-2)/4$  \\ \cline{2-9}
				& $-i$ & $0$ & $-3/8$ & $1/8$ &$-1/8$ & $-1/8$ & $(M-2)/4$ & $(M-2)/4$ \\
				& & $1/2$ & $-3/8$ & $-3/8$ & $1/8$ & $1/8$ & $(M-2)/4$  & $(M-2)/4$  \\ \hline	
				$4m+3$ & $1$ & $0$ & $3/8$ & $-3/8$ &$1/8$ & $1/8$ & $(M+1)/4$ & $(M+1)/4$ \\
				& & $1/2$ & $3/8$ & $1/8$ & $-1/8$ & $-1/8$ & $(M+1)/4$  & $(M+1)/4$  \\ \cline{2-9}
				& $i$ & $0$ & $1/8$ & $3/8$ &$-1/8$ & $-1/8$ & $(M+1)/4$ & $(M+1)/4$ \\
				& & $1/2$ & $1/8$ & $-1/8$ & $1/8$ & $1/8$ & $(M+1)/4$  & $(M+1)/4$  \\ \cline{2-9}
				 & $-1$ & $0$ & $-1/8$ & $1/8$ &$1/8$ & $1/8$ & $(M+1)/4$ & $(M+1)/4$ \\
				& & $1/2$ & $-1/8$ & $-3/8$ & $-1/8$ & $-1/8$ & $(M-3)/4$  & $(M-3)/4$  \\ \cline{2-9}
				& $-i$ & $0$ & $-3/8$ & $-1/8$ &$-1/8$ & $-1/8$ & $(M-3)/4$ & $(M-3)/4$ \\
				& & $1/2$ & $-3/8$ & $3/8$ & $1/8$ & $1/8$ & $(M+1)/4$  & $(M+1)/4$  \\ \hline	
				$4m+4$ & $1$ & $0$ & $3/8$ & $3/8$ &$1/8$ & $1/8$ & $M/4+1$ & $M/4+1$ \\
				& & $1/2$ & $3/8$ & $-1/8$ & $-1/8$ & $-1/8$ & $M/4$  & $M/4$  \\ \cline{2-9}
				& $i$ & $0$ & $1/8$ & $1/8$ &$-1/8$ & $-1/8$ & $M/4$ & $M/4$ \\
				& & $1/2$ & $1/8$ & $-3/8$ & $1/8$ & $1/8$ & $M/4$  & $M/4$  \\ \cline{2-9}
				 & $-1$ & $0$ & $-1/8$ & $-1/8$ &$1/8$ & $1/8$ & $M/4$ & $M/4$ \\
				& & $1/2$ & $-1/8$ & $3/8$ & $-1/8$ & $-1/8$ & $M/4$  & $M/4$  \\ \cline{2-9}
				& $-i$ & $0$ & $-3/8$ & $-3/8$ &$-1/8$ & $-1/8$ & $M/4-1$ & $M/4-1$ \\
				& & $1/2$ & $-3/8$ & $1/8$ & $1/8$ & $1/8$ & $M/4$  & $M/4$  \\ \hline							\end{tabular}
		}
	}
	\caption{The values of localized fluxes at fixed points and index ($l=0$) on $T^2/{\mathbb{Z}}_4$.	}
	\label{Z4tab}
\end{table}

\begin{table}[!ht]
	\centering
	{\tabcolsep = 3mm
		\renewcommand{\arraystretch}{1.2}
		\scalebox{0.80}{
			\begin{tabular}{ccc|c:c:c:c:c:c|c|c} \hline
			 flux& parity& twist & \multicolumn{6}{c|}{localized flux } & index& \eqref{Zeromodecounting}\\
				$M$ & $\eta$ & $\alpha$ & $\frac{\xi^F_{1}}{N}$& $\frac{\xi^F_{2}}{N}$ & $\frac{\xi^F_{3}}{N}$  &$\frac{\xi^F_{4}}{N}$& $\frac{\xi^F_{5}}{N}$ & $\frac{\xi^F_{6}}{N}$ &  $\frac{M}{N}+\sum_{I=1}^6 \frac{\xi^F_I}{N}$ &
				$\frac{M-V_{+}}{N}+1$\\ \hline				
				$6m+1$ & $1$ & $1/2$ & $5/12$ & $-2/12$ &$-2/12$ & $-1/12$&$-1/12$ & $-1/12$ & $(M-1)/6$ & $(M-1)/6$ \\
				 & $\omega$ & $1/2$ & $3/12$ & $2/12$ &$2/12$ & $1/12$&$1/12$ & $1/12$ & $(M+5)/{6}$ & $(M+5)/{6}$ \\
				 & $\omega^2$ & $1/2$ & $1/12$ & $0$ &$0$ & $-1/12$&$-1/12$ & $-1/12$ & $(M-1)/6$ & $(M-1)/6$ \\
				 & $\omega^3$ & $1/2$ & $-1/12$ & $-2/12$ &$-2/12$ & $1/12$&$1/12$ & $1/12$ & $(M-1)/6$ & $(M-1)/6$ \\
				 & $\omega^4$ & $1/2$ & $-3/12$ & $2/12$ &$2/12$ & $-1/12$&$-1/12$ & $-1/12$ & $(M-1)/6$ & $(M-1)/6$ \\
				 & $\omega^5$ & $1/2$ & $-5/12$ & $0$ &$0$ & $1/12$&$1/12$ & $1/12$ & $(M-1)/6$ & $(M-1)/6$ \\ \hline
				$6m+2$ & $1$ & $0$ & $5/12$ & $0$ &$0$ & $1/12$&$1/12$ & $1/12$ & $(M+4)/6$ & $(M+4)/6$ \\
				 & $\omega$ & $0$ & $3/12$ & $-2/12$ &$-2/12$ & $-1/12$&$-1/12$ & $-1/12$ & $(M-2)/6$ & $(M-2)/6$ \\
				 & $\omega^2$ & $0$ & $1/12$ & $2/12$ &$2/12$ & $1/12$&$1/12$ & $1/12$ & $(M+4)/6$ & $(M+4)/6$ \\
				 & $\omega^3$ & $0$ & $-1/12$ & $0$ &$0$ & $-1/12$&$-1/12$ & $-1/12$ & $(M-2)/6$ & $(M-2)/6$ \\
				 & $\omega^4$ & $0$ & $-3/12$ & $-2/12$ &$-2/12$ & $1/12$&$1/12$ & $1/12$ & $(M-2)/6$ & $(M-2)/6$ \\
				 & $\omega^5$ & $0$ & $-5/12$ & $2/12$ &$2/12$ & $-1/12$&$-1/12$ & $-1/12$ & $(M-2)/6$ & $(M-2)/6$ \\ \hline
				$6m+3$ & $1$ & $1/2$ & $5/12$ & $2/12$ &$2/12$ & $-1/12$&$-1/12$ & $-1/12$ & $(M+3)/6$ & $(M+3)/6$ \\
				 & $\omega$ & $1/2$ & $3/12$ & $0$ &$0$ & $1/12$&$1/12$ & $1/12$ & $(M+3)/6$ & $(M+3)/6$ \\
				 & $\omega^2$ & $1/2$ & $1/12$ & $-2/12$ &$-2/12$ & $-1/12$&$-1/12$ & $-1/12$ & $(M-3)/6$ & $(M-3)/6$ \\
				 & $\omega^3$ & $1/2$ & $-1/12$ & $2/12$ &$2/12$ & $1/12$&$1/12$ & $1/12$ & $(M+3)/6$ & $(M+3)/6$ \\
				 & $\omega^4$ & $1/2$ & $-3/12$ & $0$ &$0$ & $-1/12$&$-1/12$ & $-1/12$ & $(M-3)/6$ & $(M-3)/6$ \\
				 & $\omega^5$ & $1/2$ & $-5/12$ & $-2/12$ &$-2/12$ & $1/12$&$1/12$ & $1/12$ & $(M-3)/6$ & $(M-3)/6$ \\ \hline
				$6m+4$ & $1$ & $0$ & $5/12$ & $-2/12$ &$-2/12$ & $1/12$&$1/12$ & $1/12$ & $(M+2)/6$ & $(M+2)/6$ \\
				 & $\omega$ & $0$ & $3/12$ & $2/12$ &$2/12$ & $-1/12$&$-1/12$ & $-1/12$ & $(M+2)/6$ & $(M+2)/6$ \\
				 & $\omega^2$ & $0$ & $1/12$ & $0$ &$0$ & $1/12$&$1/12$ & $1/12$ & $(M+2)/6$ & $(M+2)/6$ \\
				 & $\omega^3$ & $0$ & $-1/12$ & $-2/12$ &$-2/12$ & $-1/12$&$-1/12$ & $-1/12$ & $(M-4)/6$ & $(M-4)/6$ \\
				 & $\omega^4$ & $0$ & $-3/12$ & $2/12$ &$2/12$ & $1/12$&$1/12$ & $1/12$ & $(M+2)/6$ & $(M+2)/6$ \\
				 & $\omega^5$ & $0$ & $-5/12$ & $0$ &$0$ & $-1/12$&$-1/12$ & $-1/12$ & $(M-4)/6$ & $(M-4)/6$ \\ \hline
				$6m+5$ & $1$ & $1/2$ & $5/12$ & $0$ &$0$ & $-1/12$&$-1/12$ & $-1/12$ & $(M+1)/6$ & $(M+1)/6$ \\
				 & $\omega$ & $1/2$ & $3/12$ & $-2/12$ &$-2/12$ & $1/12$&$1/12$ & $1/12$ & $(M+1)/6$ & $(M+1)/6$ \\
				 & $\omega^2$ & $1/2$ & $1/12$ & $2/12$ &$2/12$ & $-1/12$&$-1/12$ & $-1/12$ & $(M+1)/6$ & $(M+1)/6$ \\
				 & $\omega^3$ & $1/2$ & $-1/12$ & $0$ &$0$ & $1/12$&$1/12$ & $1/12$ & $(M+1)/6$ & $(M+1)/6$ \\
				 & $\omega^4$ & $1/2$ & $-3/12$ & $-2/12$ &$-2/12$ & $-1/12$&$-1/12$ & $-1/12$ & $(M-5)/6$ & $(M-5)/6$ \\
				 & $\omega^5$ & $1/2$ & $-5/12$ & $2/12$ &$2/12$ & $1/12$&$1/12$ & $1/12$ & $(M+1)/6$ & $(M+1)/6$ \\ \hline
				$6m+6$ & $1$ & $0$ & $5/12$ & $2/12$ &$2/12$ & $1/12$&$1/12$ & $1/12$ & $M/6+1$ & $M/6+1$ \\
				 & $\omega$ & $0$ & $3/12$ & $0$ &$0$ & $-1/12$&$-1/12$ & $-1/12$ & $M/6$ & $M/6$ \\
				 & $\omega^2$ & $0$ & $1/12$ & $-2/12$ &$-2/12$ & $1/12$&$1/12$ & $1/12$ & ${M/6}$ & $M/6$ \\
				 & $\omega^3$ & $0$ & $-1/12$ & $2/12$ &$2/12$ & $-1/12$&$-1/12$ & $-1/12$ & ${M/6}$ & $M/6$ \\
				 & $\omega^4$ & $0$ & $-3/12$ & $0$ &$0$ & $1/12$&$1/12$ & $1/12$ & ${M/6}$ & $M/6$ \\
				 & $\omega^5$ & $0$ & $-5/12$ & $-2/12$ &$-2/12$ & $-1/12$&$-1/12$ & $-1/12$ & $M/6-1$ & $M/6-1$ \\ \hline
				\end{tabular}
		}
	}
	\caption{The values of localized fluxes at fixed points and index ($l=0$) on $T^2/{\mathbb{Z}}_6$.
		}
	\label{Z6tab}
\end{table}

\renewcommand{\thesection}{\Alph{section}}

\clearpage
\bibliographystyle{unsrt}
 \bibliography{reference}

\end{document}